\documentstyle{mn}
\input{epsf}

\newif\ifAMStwofonts
\AMStwofontstrue

\def\be{\begin{equation}}
\def\ee{\end{equation}}
\def\go{\mathrel{\raise.3ex\hbox{$>$}\mkern-14mu
             \lower0.6ex\hbox{$\sim$}}}
\def\lo{\mathrel{\raise.3ex\hbox{$<$}\mkern-14mu
             \lower0.6ex\hbox{$\sim$}}}
\def\br{{\bf r}}
\def\bC{{\bf C}}
\def\bD{{\bf D}}
\def\bv{{\bf v}}
\def\bxi{{\vec \xi}}
\def\bOmega{{\bf \Omega}}
\def\Oms{\Omega_s}
\def\Omo{\Omega_{\rm orb}}
\def\omi{\omega_\alpha}
\def\omr{\omega_\alpha^{(r)}}
\def\nui{\nu_\alpha}
\def\lp{\left(}
\def\rp{\right)}
\def\lb{\left[}
\def\rb{\right]}
\def\sin{{\rm sin}}
\def\cos{{\rm cos}}
\def\lm{j}
\def\mm{k}
\def\lr{l}
\def\mr{m}
\def\mb{m'}

\title[Resonant Tidal Excitations of Rotating Neutron Stars ...]
   {Resonant Tidal Excitations of Rotating Neutron Stars in Coalescing
   Binaries}
\author[W.C.G. Ho and D. Lai]
       {Wynn C. G. Ho and Dong Lai\\
        Centre for Radiophysics and Space Research,
	Department of Astronomy, 
	Cornell University, Ithaca, NY 14853, USA\\
	{\rm E-mail: (wynnho, dong)@spacenet.tn.cornell.edu}}
\date{Accepted 1999 xxx,
      Received 1999 xxx;
      in original form 1999 xxx}

\pagerange{\pageref{firstpage}--\pageref{lastpage}}

\begin{document}

\maketitle

\label{firstpage}

\begin{abstract}
In a coalescing neutron star-neutron star or neutron star-black
hole binary, oscillation modes of the neutron star can be
resonantly excited by the companion during the final minutes of
the binary inspiral, when the orbital frequency sweeps up from a
few Hertz to a few thousand Hertz. The resulting resonant energy
transfer between the orbit and the neutron star speeds up or
slows down the inspiral, depending on whether the resonant mode
has positive or negative energy, and induces a phase change in
the emitted gravitational waves from the binary. While only g-modes
can be excited for a nonrotating neutron star, f-modes and r-modes
can also be excited when the neutron star is spinning.  A tidal
resonance, designated by the index $(jk,m)$ ($\{jk\}$ specifies
the angular order of the mode, as in the spherical harmonic $Y_{jk}$),
occurs when the mode frequency equals $m$ times the orbital frequency.
For the f-mode resonance to occur before coalescence, the neutron
star must have rapid rotation, with spin frequency $\nu_s\go 710$~Hz
for the $(22,2)$-resonance and $\nu_s\go 570$~Hz for the
$(33,3)$-resonance (assuming canonical neutron star mass,
$M=1.4~M_\odot$, and radius, $R=10$~km; however, for $R=15$~km,
these critical spin frequencies become $330$~Hz and $260$~Hz,
respectively).
Although current observations suggest that such high rotation rates
may be unlikely for coalescing binary neutron stars, these rates are
physically allowed.
Because of their strong tidal coupling, the f-mode
resonances induce a large change in the number of orbital cycles of
coalescence, $\Delta N_{\rm orb}$, with the maximum
$\Delta N_{\rm orb}\sim 10-1000$ for the $(22,2)$-resonance
and $\Delta N_{\rm orb} \sim 1$ for the $(33,3)$-resonance. 
Such f-mode resonant effects, if present, must be included in constructing
the templates of waveforms used in searching for gravitational wave signals.
Higher order f-mode resonances can occur at slower rotation rates, 
but the induced orbital change is much smaller ($\Delta N_{\rm orb}\lo 0.1$). 
For the dominant g-mode $(22,2)$-resonance, even modest rotation
($\nu_s\lo 100$~Hz) can enhance the resonant effect on the orbit
by shifting the resonance to a smaller orbital frequency.  However,
because of the weak coupling between the g-mode and the tidal potential,
$\Delta N_{\rm orb}$ lies in the range $10^{-3}-10^{-2}$ (depending
strongly on the neutron star equation of state) and is probably
negligible for the purpose of detecting gravitational waves.  Resonant
excitations of r-modes require misaligned spin-orbit inclinations, and
the dominant resonances correspond to octopolar excitations of the
$\lm=\mm=2$ mode, with $(\lm\mm,\mr) = (22,3)$ and $(22,1)$. 
Since the tidal coupling of the r-mode depends strongly on rotation rate, 
$\Delta N_{\rm orb}\lo 10^{-2}(R/10\,{\rm km})^{10}
(M/1.4\,M_\odot)^{-20/3}$ is negligible for canonical neutron star
parameters but can be appreciable if the neutron star radius is larger.
\end{abstract}


\section{Introduction}
The new generation of gravitational wave detectors such as 
the Laser Interferometer Gravitational Wave Observatory (LIGO;
Abramovici et al.~1992) and its French-Italian
counterpart VIRGO (Bradaschia et al.~1990) are
designed to observe gravitational radiation in the frequency
range from $\sim$~10--1000~Hz.  Coalescing neutron star-neutron star
(NS-NS) and neutron star-black hole (NS-BH) binaries are promising
sources for such radiation, with the last few minutes of the
binary inspiral producing gravitational waves which sweep upward
through this frequency range (e.g., Thorne~1987, Cutler et al.~1993,
Thorne~1998).  Due to the low signal-to-noise ratios, a good
understanding of the mechanisms that generate and affect the
gravitational waveform is required to create accurate theoretical
templates that can be used to extract the signals from the noise.

The binary inspiral and the resulting waveform can be described, to
leading order, by Newtonian dynamics of two point masses, together
with the lowest order dissipative effect corresponding to the
emission of gravitational radiation via the quadrupole formula.
To more accurately construct the theoretical templates, general
relativistic effects must be taken into account by using 
post-Newtonian expansions (e.g., Blanchet et al.~1995) or techniques of
numerical relativity (e.g., Teukolsky 1998; Brady et al.~1998).
Other corrections come from hydrodynamical effects 
due to the finite size of neutron stars (Kochanek 1992; Bildsten \& Cutler
1992; Lai et al.~1994). In particular, tidal interactions
can have significant effects on the final stage of binary 
inspiral by destabilizing the orbit and accelerating the coalescence
at small orbital radii (e.g., Lai et al.~1994; Lai \& Shapiro~1995;
Rasio \& Shapiro~1995; Lai \& Wiseman~1996;
Baumgarte et al.~1997; New \& Tohline~1997;
Uryu \& Eriguchi~1998; see also Shibata et al.~1993; Davies et al.~1994; 
Zhuge et al.~1996; Ruffert et al.~1997 for numerical
simulations on binary coalescence).

The early stage of the inspiral, with the gravitational wave frequencies
in the range of $10$~Hz to a few hundred Hz, is particularly important
for detecting the wave signal and for extracting binary parameters
from the waveform using the matched filtering technique (e.g., Cutler et
al.~1993; Cutler \& Flanagan 1994). It is usually thought that tidal effects
are completely negligible for this ``low-frequency'' regime. 
This is indeed the case for the ``quasi-equilibrium'' tides
as the tidal interaction potential scales as
$D^{-6}$ (where $D$ is the orbital separation; see
Lai et al.~1994 for analytic expressions of the orbital phase error
induced by static tides). The situation is more
complicated for the {\it resonant tides}: 
As two compact objects inspiral, the orbit can momentarily
come in resonance with the normal oscillation modes of the NS.
By drawing energy from the orbital motion and resonantly exciting
the modes, the rate of inspiral is modified, giving rise to a
phase error in the gravitational waveform.  
This situation has been studied (Reisenegger \& Goldreich 1994; Lai 1994)
in the case of nonrotating neutron stars where the only modes
that can be resonantly excited are g-modes (with typical mode
frequencies $\lo 100$~Hz). It was found that the effect is negligible because
the coupling between the g-mode and the tidal potential is weak.
However, the NS can be rotating, and consequently, these results need to 
be reexamined. As we show in this paper, even modest rotation of the NS
can enhance the resonant g-mode excitation by shifting 
the resonance to a lower frequency (where the orbital decay is slower
and resonance is stronger). In addition, f-modes, which possess 
strong tidal coupling, can have their normally high frequencies in the
nonrotating case ``dragged'' or reduced by retrograde rotation to 
sufficiently low frequencies in the inertial frame, such that resonant
excitations can occur well before binary merger. The NS spin also 
gives rise to a new class of modes which are absent in the nonrotating case:
R-modes, which can have low frequencies and may also undergo resonant
excitations.  

The purpose of this paper is to study the resonant excitations
of various modes of spinning NSs in inspiraling binary systems
and to examine their effects on the orbital decay rate
(particularly the phase error in the gravitational waveform).
Section 2 develops the general formalism of resonant tides in
coalescing NS binaries. In section 3 we apply the analysis to
f-modes, g-modes, and r-modes when the spin axis is aligned or
anti-aligned with the orbital angular momentum axis. Section 4
describes the modification to the results of the aligned
(anti-aligned) case when the spin and orbital axes are
misaligned by an inclination angle $\beta$.  In section 5, we
summarize our results and discuss their implications for
gravitational wave detection.

\section{Resonant Mode Excitation}

\subsection{Basic equations}

Consider a NS with mass $M$, radius $R$, and spin $\bOmega_s$ in
orbit with a companion $M'$, which can be another NS or a BH.  We
treat $M'$ as a point mass.  The orbital frequency vector is
$\bOmega_{\rm orb}$.  
We define a coordinate system $(x,y,z)$ centered on $M$ with
the $z$-axis along $\bOmega_{\rm orb}$, such that
$\bOmega_{\rm orb} = \Omo{\hat z}$ (with $\Omo\ge 0$).
Define another coordinate system $(x',y',z')$ with
the $z'$-axis along $\bOmega_s$, such that
$\bOmega_s = \Oms{\hat z'}$ (with $\Oms\ge 0$)
and the $y'$-axis in the orbital plane; these are the body axes.
Let the angle between the $z$-axis and $z'$-axis be $\beta$
(the spin-orbit inclination angle) and the angle between
the $y$-axis and $y'$-axis be $\alpha$. Thus the $(x',y',z')$ frame is
related to the $(x,y,z)$ frame by Euler angle $(\alpha,\beta,\gamma=0)$.

The gravitational potential produced by $M'$ can be expanded in terms of
spherical harmonics: 
\begin{eqnarray}
U(\br,t) & = & -{GM'\over |\br-\bD(t)|} \nonumber \\
 & = & -GM'\sum_{\lr\mr}{W_{\lr\mr}r^\lr
\over [D(t)]^{\lr+1}} e^{-i\mr\Phi(t)}
Y_{\lr\mr}(\theta,\phi), \label{eq:potential}
\end{eqnarray}
where $\br=(r,\theta,\phi)$ is the position vector of a fluid element 
in star $M$ ($r,\theta,\phi$ are the spherical coordinates with respect to the
$z$-axis), $\bD(t)=[D(t),\pi/2,\Phi(t)]$ is the position vector 
of the point mass, $M'$, and $W_{lm}$ is defined in 
Press \& Teukolsky (1977) as
\begin{eqnarray}
W_{lm} & = & (-)^{(l+m)/2}\lb{4\pi\over 2l+1}(l+m)!(l-m)!\rb^{1/2}
\nonumber \\
&& \times \lb2^l\lp{l+m\over 2}\rp!\lp{l-m\over 2}\rp!\rb^{-1}.
\label{eq:wlm}
\end{eqnarray}
Here the symbol $(-)^p$ is zero if $p$ is not an integer.
For example, $W_{2\pm 2} = (3\pi/10)^{1/2}$ and
$W_{3\pm 3} = (5\pi/28)^{1/2}$.  
In equation (\ref{eq:potential}), the $\lr=0$ and $\lr=1$ terms can be
dropped since they are not relevant for tidal deformation.  The
tidal potential is specified by the index $\{lm\}$ (relative to the
orbital angular momentum axis $\hat z$) and is nonzero only when
$(\lr + \mr)$ is even, reflecting the symmetry of the tidal potential.

It is more convenient to define oscillation modes relative to
the spin axis $\hat z'$, and we will need to express
the tidal potential in terms of $Y_{lm'}(\theta',\phi')$,
the spherical harmonic function relative to
$\hat z'$. The function $Y_{\lr\mr}(\theta,\phi)$ is related to
$Y_{\lr\mb}(\theta',\phi')$ by
\be
Y_{\lr\mr}(\theta,\phi)=\sum_{m'}{\cal D}^{(\lr)}_{\mb\mr}(\alpha,\beta)
Y_{\lr\mb}(\theta',\phi'),
\ee
where the Wigner ${\cal D}$-function is given by (e.g., Wybourne 1974)
\begin{displaymath}
{\cal D}^{(\lr)}_{\mb\mr}(\alpha,\beta) = e^{i\mb\alpha}\!
\lb(\lr+\mr)!\,(\lr-\mr)!\,(\lr+\mb)!\,(\lr-\mb)!\rb^{1/2}
\end{displaymath}
\be
\times \sum_p{(-1)^{\lr+\mb-p}\lp\cos{\beta\over 2}\rp^{2p-\mr-\mb}\!\!
\lp\sin{\beta\over 2}\rp^{2\lr-2p+\mr+\mb}
\over p!\,(\lr+\mr-p)!\,(\lr+\mb-p)!\,(p-\mr-\mb)!}.
\label{eq:wigd}
\ee
Thus $\{lm'\}$ is the tidal potential index relative to the spin axis. 

The linear perturbation of the tidal potential on $M$ is specified
by the Lagrangian displacement, $\bxi(\br,t)$, of a fluid element
from its unperturbed position.  In the inertial frame, the
equation of motion takes the form
\be
\frac{\partial^2 \bxi}{\partial t^2}+2(\bv\cdot\nabla)
\frac{\partial\bxi}{\partial t}+{\bC}\cdot\bxi=-\nabla U, \label{eq:eqnmotion2}
\ee
where $\bv = \bOmega_s \times \br$ is the unperturbed fluid velocity
and $\bC$ is a self-adjoint operator (a
function of the pressure and gravity perturbations)
acting on $\bxi$ (Friedman \& Schutz 1978a).
Our analysis is accomplished through an eigenmode decomposition
of the Lagrangian displacement:  $\bxi$ is taken to be the
linear superposition of eigenmodes, $\bxi_\alpha(\br,t)$, namely
\be
\bxi(\br,t)=\sum_\alpha a_\alpha(t)\bxi_\alpha(\br). \label{eq:ximode1}
\ee
We shall specify the normal mode by the index $\alpha=\{njk\}$ with respect
to the body axes: $n$ is the number of radial nodes, $j$ gives the
number of nodes in the $\theta'$-eigenfunction, and $k$ is the azimuthal
index (about the $z'$-axis). In the following we shall suppress the radial
index $n$ so that $\alpha=\{jk\}$. In the absence of external perturbation,
the eigenmode behaves as 
\be
\bxi_\alpha(\br,t)=\bxi_\alpha(\br)e^{i\omi t} \propto e^{i\omi t+ik\phi'},
\label{eq:ximode2}
\ee
where $\omi$ is the $\alpha$-mode frequency in the inertial frame.
The mode eigenfunction satisfies the equation
\be
-\omi^2\bxi_\alpha+2i\omi(\bv\cdot\nabla)\bxi_\alpha
+\bC\cdot\bxi_\alpha=0, \label{eq:eqnmotion1}
\ee
and we shall adopt the normalization
\be
\int\!d^3\!x\,\rho\,\bxi_\alpha^\ast\cdot\bxi_{\alpha'}=\delta_{\alpha\alpha'}.
\label{eq:xinorm}
\ee

Substituting (\ref{eq:ximode1}) into (\ref{eq:eqnmotion2}), using
(\ref{eq:ximode2}) and (\ref{eq:xinorm}), and noting that
(\ref{eq:eqnmotion1}) gives a relation for $\bC\cdot\bxi$, the
dynamical equation for the mode amplitude is
\begin{displaymath}
\ddot a_\alpha -2iT_\alpha\dot a_\alpha + (\omi^2 - 2T_\alpha \omi)
a_\alpha
\end{displaymath}
\be
\qquad \qquad = \sum_{\lr\mr\mb} \frac{GM'}{D^{\lr+1}}
W_{\lr\mr} {\cal D}^{(\lr)}_{\mb\mr} Q_{\alpha,\lr\mb}
e^{-i\mr\Phi}, \label{eq:eqnmotion3}
\ee
where
\begin{eqnarray}
T_\alpha &=& i\!\!\int\!d^3\!x\,\rho\,\bxi_\alpha^\ast\cdot(\bv\cdot\nabla)
\bxi_\alpha\nonumber\\
&=&-k\Omega_s+i\int\!d^3\!x\,\rho\,\bxi_\alpha^\ast\cdot
({\bf \Omega}_s\times\bxi_\alpha)
\label{eq:t}\end{eqnarray}
(this is called $B_\alpha$ in Lai 1997)
and $Q_{\alpha,\lr\mb}$ is the tidal coupling coefficient or overlap integral
defined by
\begin{eqnarray}
Q_{\alpha,\lr\mb} & = & \int\!d^3\!x'\,\rho\,{\bxi}_\alpha^\ast\cdot
\nabla\lb r^\lr Y_{\lr\mb}(\theta',\phi')\rb \nonumber \\
 & = & \int\!d^3\!x'\,\delta\rho_\alpha^\ast
\lb r^\lr Y_{\lr\mb}(\theta',\phi')\rb. \label{eq:qcoup}
\end{eqnarray}
Here $\delta\rho_\alpha=-\nabla\cdot(\rho\bxi_\alpha)$ is the
Eulerian density perturbation.
Clearly, $Q_{\alpha,lm'}$ is nonzero only when $m'=k$, thus
\be
Q_{\alpha,lm'}=Q_{jk,lk}\delta_{m'k}.
\ee

\subsection{Resonant Mode Energy}

The right-hand-side of equation (\ref{eq:eqnmotion3}) contains 
a sum of driving terms with time dependence proportional to
$e^{-im\Phi(t)}$. For a given mode $\alpha=\{jk\}$, resonance occurs when
\be
|\omi|=|m|\Omega_{\rm orb},~~~~~({\rm Resonance~Condition})
\label{con1}\ee
provided there exists at least one $l$ (obviously $l \ge |m|,~l\ge|k|$)
which satisfies 
\be
l+m={\rm even}~~~~~({\rm Symmetry~of~the~Tidal~Potential})
\label{con2}\ee
and 
\be
{\cal D}_{km}^{(l)}Q_{jk,lk}\ne 0.~~~~~({\rm Nonzero~Tidal~Coupling})
\label{con3}\ee
For $Q_{\lm\mm,\lr\mm}$ to be nonzero, the projected tidal
potential [$\propto Y_{\lr\mm}(\theta',\phi')$] must have
the same $\theta'$-symmetry, with respect to the equator, as
the mode.  This gives the requirement that $\lr+\lm=$ even (for
spherical modes) or $\lr+\lm=$ odd (for r-modes) (see \S 3). 

We consider now a specific resonance. 
Hereafter we shall adopt the convention\footnote{Clearly,
in equations (\ref{eq:ximode2}) and (\ref{eq:eqnmotion3}),
$k$ and $m$ can be positive or negative.}
$k>0$ and $m>0$. Thus a resonance is designated by 
the indices $(\alpha,m)=(jk,m)$, which indicates that the mode
$\alpha=\{jk\}$ is excited by tidal potential of azimuthal order $m$
(with angular dependence $e^{\pm im\Phi}$).
In general, there can be many $l$'s that satisfies
the conditions (\ref{con1})-(\ref{con3})\footnote{The exception
is when the star is nonrotating, in which case $m=k$ and $l=j$.},
but clearly the smallest 
$l$ which satisfies these conditions will provide the dominant 
contribution. When appropriate, we shall also label a resonance using the
notation $(jk,lm)$, with $l$ specifying the angular order
of the dominant tidal field that contributes the resonance
$(jk,m)$.

In the special case when the spin and orbit are aligned or anti-aligned,
one must have $m=k$, i.e., each mode can experience resonance only once. 
But in the general case of arbitrary spin-orbit inclination angle, $\beta$,
for a given mode $\{jk\}$, there may be infinite number of $m$ which 
satisfies the conditions (\ref{con1})-(\ref{con3});
in other words, the mode can be
resonantly excited many times at different orbital frequencies. Thus, 
strictly speaking, the energy transfer from the orbit to the mode
at a given resonance $(jk,m)$ depends on the amplitude of the mode
induced by previous resonances $(jk,m+2),~(jk,m+4),\cdots$
\footnote{Consider a spheroidal (f or g) mode, $\alpha = \{\lm\mm\}$.
For the $(\lm\mm,\mr)$-resonance to be in effect, it is necessary
for $\lr+\mr =$ even [see eq.~(\ref{con2})] and $\lr+\lm =$ even
(otherwise $Q_{\lm\mm,\lr\mm} = 0$).  Now in order for the
$(\lm\mm,\mr+1)$ to be in effect, it is necessary for $\lr+\mr+1=$ even,
which is incompatible with $\lr+\lm=$ even.}.
However, as we shall see below, in almost all cases of interest, the
energy transfer at the $(jk,m)$-resonance, $\Delta E_{jk,m}$,
is much greater than the energy transfer at the earlier resonances,
i.e., $\Delta E_{jk,m}\gg \Delta E_{jk,m+2}$. Therefore, throughout
this paper, we shall treat different resonances of the same mode
as being independent of each other. This corresponds to assuming
that before each resonance, the mode amplitude is essentially zero. 

To calculate the energy transfer to a given mode at a particular
$\mr$-resonance, $\Delta E_{jk,m}$, we only need to keep the 
resonant driving terms in (\ref{eq:eqnmotion3}), namely
\begin{displaymath}
\ddot a_\alpha -2iT_\alpha\dot a_\alpha + (\omi^2 - 2T_\alpha \omi)
a_\alpha
\end{displaymath}
\begin{displaymath}
\quad \qquad = \lp\sum_{\lr} \frac{GM'}{D^{\lr+1}}
W_{\lr\mr} {\cal D}^{(\lr)}_{\mm\mr} Q_{\alpha,\lr\mm}\rp
e^{-i\mr\Phi}
\end{displaymath}
\be
\qquad \qquad + \lp\sum_{\lr} \frac{GM'}{D^{\lr+1}}
W_{\lr\mbox{-}\mr} {\cal D}^{(\lr)}_{\mm\mbox{-}\mr} Q_{\alpha,\lr\mm}\rp
e^{i\mr\Phi}. \label{eq:eqnmotion4}
\ee
Since $a_\alpha\propto e^{i\omi t}$ in the absence of external forcing
[see eq.~(\ref{eq:ximode2})], it is natural to define the mode amplitude
$c_\alpha$ via
\be
a_\alpha = c_\alpha e^{i\omi t}, \label{eq:xiamp}
\ee
and we expect $c_\alpha$ to be independent of time after the resonance
\footnote{Before and after a resonance, there is a small, non-resonant 
contribution to $a_\alpha$ which is proportional to $e^{\pm im\Phi(t)}$.
This contribution can be neglected compared to the resonance-induced 
amplitude.}.
Equation (\ref{eq:eqnmotion4}) then becomes
\begin{displaymath}
\ddot c_\alpha +2i(\omi - T_\alpha)\dot c_\alpha
\end{displaymath}
\begin{displaymath}
\quad \qquad = \lp\sum_{\lr} \frac{GM'}{D^{\lr+1}}
W_{\lr\mr} {\cal D}^{(\lr)}_{\mm\mr} Q_{\alpha,\lr\mm}\rp
e^{-i\mr\Phi-i\omi t}
\end{displaymath}
\be
\qquad \qquad + \lp\sum_{\lr} \frac{GM'}{D^{\lr+1}}
W_{\lr\mbox{-}\mr} {\cal D}^{(\lr)}_{\mm\mbox{-}\mr} Q_{\alpha,\lr\mm}\rp
e^{i\mr\Phi-i\omi t}.
\label{eq:eqnmotion5}
\ee
Only one of the terms on the RHS of (\ref{eq:eqnmotion5})
can induce the resonance, which occurs when 
the argument of the exponential of either the first term or the
second term goes through zero.  The term that induces resonance
depends on the sign of $\omi$. 

\subsubsection{General Properties of the Mode Frequencies}

Before proceeding with our calculation, let us briefly discuss
various possibilities for the mode frequency. In general, 
for a given mode index $\alpha=\{jk\}$ (recall our sign convention
$k>0$, and note that we have suppressed the radial mode index $n$),
there are two possible solutions of opposite signs
for $\omr$, the mode frequency in the frame
corotating with the star, which correspond to two different directions of
propagation for the wave pattern. Recall from (\ref{eq:ximode2}) 
that our convention for free oscillations is $e^{i\omi t+ik\phi'}$ relative to
the spin axis, where ${\bOmega_s} = \Oms{\hat z}'$ and $\Oms\ge 0$.
The mode frequency in the inertial 
frame, $\omi$, is related to $\omr$ by
\be
\omi = \omr - k \Oms. \label{eq:omegar}
\ee

The first solution, with $\omr>0$, corresponds to the mode
which propagates retrograde with respect to the spin (in
the rotating frame). Rotation drags the mode frequency in the
inertial frame to small values; thus the resonance is shifted to smaller
orbital frequency. With increasing $\Oms$, the mode frequency $\omi$ 
may become negative, corresponding to wave propagation in the
same direction as the spin (in the inertial frame).  This
change in sign of $\omi$ signals the onset of
the so-called Chandrasekhar-Friedman-Schutz (CFS) instability
(Chandrasekhar 1970; Friedman \& Schutz 1978b) and can occur
for both f-modes (e.g., Lindblom 1995) and g-modes (Lai 1998). 
At this point the mode energy in the inertial frame
changes from positive (stable mode) to negative (unstable
mode)\footnote{Since the growth time of the instability
can be very long, and since viscosity can suppress the CFS 
instability for some modes (e.g., Lindblom 1995), we shall
still consider such unstable modes
in this paper.}. Note that when the resonant excitation of a positive-energy
mode occurs, the orbital energy is transferred to the mode and the orbit 
decays faster; when a negative-energy mode is excited, the orbit actually 
gains energy and the orbital decay is slowed. 

The second solution, with $\omr < 0$, propagates prograde with
respect to the spin (in the rotating frame).
Rotation causes $\omi$ to become even more negative. This mode always has
positive energy and is stable. The resonance is shifted to even
higher orbital frequency. 
Thus for f-modes, the $\omr<0$ modes do not experience any resonance.
For g-modes, $|\omi|$ can still lie in the appropriate range for
resonance.

The r-modes are a special case for which $\omr>0$ but $\omi<0$,
meaning that in the rotating frame, the r-mode wave pattern always
propagates opposite to the spin, while in the inertial frame it
propagates in the same direction as the spin. This implies that
the r-modes are always CFS unstable in the absence of fluid
viscosity (see \S 3.3 below). 

\subsubsection{Mode with $\omi>0$ (Spin-Retrograde Mode)}

For a mode with $\omi>0$, which corresponds to wave propagation 
opposite to the rotation, resonance occurs at 
$m\dot\Phi = m\Omo = \omi$, where $\Omo$ is given by
\be
\Omo^2=\frac{G(M+M')}{D^3},
\ee
neglecting tidal and post-Newtonian corrections.
In this case, only the second term on the RHS of (\ref{eq:eqnmotion5})
can induce resonance. Thus we have
\begin{displaymath}
\ddot c_\alpha +2i(\omi - T_\alpha)\dot c_\alpha
\end{displaymath}
\be
\quad \qquad = \lp\sum_l \frac{GM'}{D^{\lr+1}} W_{\lr\mbox{-}\mr}
{\cal D}^{(\lr)}_{\mm\mbox{-}\mr} Q_{\alpha,\lr\mm} \rp
e^{i\mr\Phi - i\omi t}.
\ee
Integrating over time around the resonance and noting that
$c_\alpha$ is almost constant after the resonance 
(and $c_\alpha=0$ before the resonance since the mode amplitude 
induced by previous resonances is negligible; see the discussion at the
beginning of \S 2), we find 
\begin{displaymath}
c_\alpha \mbox{(after resonance)} = \frac{1}{2i(\omi - T_\alpha)}
\end{displaymath}
\be
\quad \times \int\!dt\,\lp\sum_l\frac{GM'}
{D^{\lr+1}} W_{\lr\mbox{-}\mr} {\cal D}^{(\lr)}_{\mm\mbox{-}\mr}
Q_{\alpha,\lr\mm}\rp e^{i\lp\mr\Phi - \omi t\rp}.
\ee
Resonance occurs when the argument of the exponential is equal to
zero.  Doing a Taylor expansion of the argument of the
exponential around the resonance epoch and using $m\dot\Phi=\omi$, we have
\be
\mr\Phi - \omi t = \mr\dot\Phi t + \mr\frac{1}{2} \ddot\Phi t^2 - \omi t
= \frac{\mr}{2} \dot\Omo t^2.
\ee
$D$ varies slowly with time, so that
\be
\int\!dt\,\frac{1}{D^{\lr+1}}e^{-\frac{1}{2}i\mr\dot\Omo t^2}
\simeq \frac{1}{D^{\lr+1}} \lp\frac{2\pi}{\mr\dot\Omo}\rp^{1/2}.
\ee
Therefore the mode amplitude after the $(\alpha,m)$-resonance is
\begin{displaymath}
a_{\alpha}(t) = \frac{e^{i\omi t}}{2i(\omi - T_\alpha)}
\end{displaymath}
\be
\quad \qquad \times \lp\sum_l \frac{GM'}{D_\alpha^{\lr+1}}
W_{\lr\mbox{-}\mr} {\cal D}^{(\lr)}_{\mm\mbox{-}\mr} Q_{\alpha,\lr\mm}\rp
\lp\frac{2\pi}{\mr\dot\Omo}\rp_{\!\alpha}^{\!\!1/2}\!\!\!,
\ee
where $D_\alpha$ is the orbital radius where resonance occurs and
$\dot\Omo$ is evaluated at $D_\alpha$. 
The energy of the mode in the inertial frame 
is given by Friedman \& Schutz (1978a) as
\be
E = \frac{1}{2} \int\!d^3\!x\,\rho\,\lp
\left|\frac{\partial\bxi}{\partial t}\right|^2
+ \bxi^\ast\cdot\bC\cdot\bxi\rp. \label{eq:energyi}
\ee
Using eqs.~(\ref{eq:ximode1})-(\ref{eq:xinorm}), this gives
the mode energy associated with the $(\alpha,m)$-resonance as
\begin{eqnarray}
\Delta E_{\alpha,m} &=& 2\lb\frac{1}{2}|\dot a_{\alpha}|^2
+ \frac{1}{2}(\omi^2-2T_\alpha\omi)|
a_{\alpha}|^2\rb \nonumber\\
&=& 2\omi(\omi-T_\alpha)|a_{\alpha}|^2 \nonumber\\
&=& \lp\!\frac{\pi}{\mr\dot\Omo}\!\rp_{\!\!\alpha}\!\!
\lp\frac{\omi}{\omi - T_\alpha}\rp \nonumber\\
&& \times \left(\sum_l \frac{GM'}{D_\alpha^{\lr+1}}
W_{\lr\mbox{-}\mr} {\cal D}^{(\lr)}_{\mm\mbox{-}\mr} Q_{\alpha,\lr\mm}
\right)^2.
\label{eq:modeen3}
\end{eqnarray}
The factor of two in the first line of (\ref{eq:modeen3})
arises because, for each $\mm > 0$ and $\omi > 0$ resonant
mode, there is also the identical $\mm < 0$ and $\omi < 0$
resonant mode (recall our sign convention $\mr>0$ and $\mm>0$).
Hereafter we shall define $Q_{\alpha,lk}$ in units such that
$G=M=R=1$. Equation (\ref{eq:modeen3}) then becomes
\begin{eqnarray}
\Delta E_{\alpha,m} &=& \lp\frac{GM}{R^3}\rp \lp\frac{GM'^2}{R}\rp
\lp\frac{\pi}{\mr\dot\Omo}\rp_{\!\!\alpha}
\frac{\omi}{\varepsilon_\alpha}\nonumber\\
&&\times \left[\sum_l W_{\lr\mbox{-}\mr}
{\cal D}^{(\lr)}_{\mm\mbox{-}\mr}Q_{\alpha,\lr\mm}
\lp\frac{R}{D_\alpha}\rp^{l+1}\right]^2.
\label{eq:fenergy}
\end{eqnarray}
Here we have defined 
\be
\varepsilon_\alpha = \omi - T_\alpha=\omr
-i\int\!d^3\!x\,\rho\,\bxi_\alpha^\ast\cdot
({\bf \Omega}_s\times\bxi_\alpha).
\label{eq:epsilon}\ee
Since the effect of the tidal interaction on the orbit is small, we
use the quadrupole formula for the gravitational radiation of two
point masses to obtain the orbital decay rate
(e.g., Shapiro \& Teukolsky 1983) 
\be
\dot\Omo=-\frac{3}{2}\frac{\dot D}{D}\Omo,
~~~~\dot D=-{64G^3\over 5c^5}{M^3q(1+q)\over D^3},
\ee
where $q=M'/M$ is the mass ratio. Equation (\ref{eq:fenergy})
can be expressed in terms of index $(\alpha,m)=(jk,m)$ and
quantities associated with the resonant mode, giving
\begin{displaymath}
\Delta E_{\alpha,\mr} = \frac{5\pi}{96} \lp\frac{GM^2}{R}\rp
\lp\frac{Rc^2}{GM}\rp^{\!5/2} \frac{q}{\hat \varepsilon_\alpha}
\end{displaymath}
\be
\times \left[\sum_l\frac{1}{(1+q)^{(2l+1)/6}}
\lp\!\frac{\hat \omi}{\mr}\!\rp^{\!\!2(\lr-1)/3}\!\!\!\!
W_{\lr\mbox{-}\mr}{\cal D}^{(\lr)}_{\mm\mbox{-}\mr}Q_{\alpha,\lr\mm}\right]^2
\!\!\!\!,
\label{energy1}
\ee
where we have defined dimensionless quantities
\be
\hat \omi = \omi\left({R^3\over GM}\right)^{1/2},
\ee
\be
\hat \varepsilon_\alpha = \varepsilon_\alpha\lp{R^3\over GM}\rp^{1/2},
\ee
\be
\hat T_\alpha = T_\alpha\left({R^3\over GM}\right)^{1/2}.
\ee
In this case, the mode energy is always positive (see \S 2.2.1).

\subsubsection{Mode with $\omi<0$ (Spin-Prograde Mode)}

For a mode with $\omi<0$, corresponding to wave propagation 
in the same direction as the spin, the resonance occurs at 
$m\Omo = -\omi$, and only the first term on the RHS of
eq.~(\ref{eq:eqnmotion5}) can induce resonance. Carrying 
out a similar calculation as in the $\omi>0$ case, we find
\begin{displaymath}
\Delta E_{\alpha,\mr} = -\frac{5\pi}{96} \lp\frac{GM^2}{R}\rp
\lp\frac{Rc^2}{GM}\rp^{\!5/2} \frac{q}{\hat \varepsilon_\alpha}
\end{displaymath}
\be
\times\left[\sum_l\frac{1}{(1+q)^{(2l+1)/6}}\!
\lp\!\frac{|\hat \omi|}{\mr}\!\rp^{\!\!2(\lr-1)/3}\!\!\!\!\!
W_{\lr\mr} {\cal D}^{(\lr)}_{km} Q_{\alpha,\lr\mm}\right]^2\!\!.
\label{energy2}
\ee
Note that the mode energy can be either positive or negative, 
depending on the sign of $\omi\varepsilon_\alpha$
[see \S 2.2.1 and eq.~(\ref{eq:modeen3})]. 
 
Apart from the sign in $\omi$, the only difference between
(\ref{energy1}) and (\ref{energy2}) is the factor ${\cal D}_{k-m}^{(l)}$
and ${\cal D}_{km}^{(l)}$. This difference reflects the 
``preferred'' spin-orbit orientation for resonant excitation: 
In order to have resonance, the wave pattern of the mode 
in the inertial frame must propagate in the same direction
as the orbital motion of the external mass, $M'$. 
For $\omi>0$, the wave propagates opposite to
the spin, thus the resonance will be strongest for
retrograde spin-orbit inclination ($\beta>90^\circ$);
For $\omi<0$, the wave propagates in the same sense as
the spin, thus we expect the resonance to be strongest for
prograde spin-orbit inclination ($\beta<90^\circ$).

\subsection{Change in Orbital Phase}

We now calculate the orbital phase error due to the resonant 
energy transfer between the binary orbit and the NS oscillation mode. 
The number of orbital cycles, $N_{\rm orb}$, is obtained from
\be
dN_{\rm orb}={\Omo\over 2\pi}dt={\Omo\over 2\pi}{dE_{\rm tot}\over 
\dot E_{\rm tot}},
\ee
where $E_{\rm tot}=E_{\rm orb}+E_{\rm star}$ is the total energy of
the system, including both the orbit and the stellar mode.
Comparing to the fiducial situation in which the mode energy is neglected
and $dN_{\rm orb}^{(0)}=(\Omo/2\pi)(dE_{\rm orb}/\dot E_{\rm tot})$,
we find that the change in the number of orbital cycles due to 
the $(\alpha,m)$-resonance is (Lai 1994)
\begin{eqnarray}
(\Delta N_{\rm orb})_{\alpha,m} & = & -\left[{t_D\over t_{\rm orb}}
\frac{\Delta E_{\alpha,\mr}}{|E_{\rm orb}|}\right]_\alpha,
\end{eqnarray}
where $t_D\equiv |E_{\rm orb}/\dot E_{\rm tot}|\simeq 
D/|\dot D|$ is the orbital decay time,
$t_{\rm orb} = 2\pi/\Omo$ is the orbital period,
and $|E_{\rm orb}| = GMM'/(2D)$ is the orbital energy.
The subscript $\alpha$ implies that $t_D,~t_{\rm orb}$, and
$|E_{\rm orb}|$ are evaluated at the resonance radius $D_\alpha$. 
Parameterizing by the resonance index $(\alpha,\mr)=(jk,m)$ and substituting
in the expressions for $\Delta E_{\alpha,m}$, we obtain,
for the mode with $\omi>0$,
\begin{displaymath}
(\Delta N_{\rm orb})_{\alpha,m} = -\frac{5}{64\pi} \lp\frac{R}{GM^2}\rp
\lp\frac{Rc^2}{GM}\rp^{5/2}
\end{displaymath}
\be
\qquad \qquad \qquad \times \frac{(1+q)^{2/3}}{q^2}
\lp\frac{\mr}{\hat \omi}\rp^{7/3} \Delta E_{\alpha,\mr}
\ee
\begin{displaymath}
\, \quad \qquad \qquad = -\frac{25}{6144} \lp\frac{Rc^2}{GM}\rp^5
\frac{1}{q\,\hat \varepsilon_\alpha}
\end{displaymath}
\be
\qquad \qquad \times\left[\sum_l
\frac{W_{\lr\mbox{-}\mr}{\cal D}^{(\lr)}_{\mm\mbox{-}\mr}
Q_{\alpha,\lr\mm}}{(1+q)^{(2\lr-1)/6}} 
\lp\frac{\hat \omi}{\mr}\rp^{\!(4\lr-11)/6}
\right]^2. \label{eq:orbchange1}
\ee
If we keep only the dominant $l$-term and label the resonance by
$(\alpha,lm)$, we arrive at
\begin{displaymath}
(\Delta N_{\rm orb})_{\alpha,lm} 
= -\frac{25}{6144} \lp\frac{Rc^2}{GM}\rp^5
\frac{1}{q(1+q)^{(2\lr-1)/3}}
\end{displaymath}
\be
\qquad \qquad \times
\frac{1}{\hat \varepsilon_\alpha}
\lp\frac{\hat \omi}{\mr}\rp^{(4\lr-11)/3}
W_{\lr\mbox{-}\mr}^2 |{\cal D}^{(\lr)}_{\mm\mbox{-}\mr}|^2
|Q_{\alpha,\lr\mm}|^2. \label{eq:orbchange}
\ee
Note that for $\omi>0$, the mode energy is always positive, and thus
$(\Delta N_{\rm orb})_{\alpha,lm}$ is negative, which implies that
the resonance causes the orbit to decay faster.

Similarly, for the mode with $\omi<0$, we find 
\begin{displaymath}
(\Delta N_{\rm orb})_{\alpha,lm} = +\frac{25}{6144} \lp\frac{Rc^2}{GM}\rp^5
\frac{1}{q(1+q)^{(2\lr-1)/3}}
\end{displaymath}
\be
\qquad \qquad \times
\frac{1}{\hat \varepsilon_\alpha}
\lp\frac{|\hat \omi|}{\mr}\rp^{\!\!(4\lr-11)/3}
W_{\lr\mr}^2 |{\cal D}^{(\lr)}_{\mm\mr}|^2 |Q_{\alpha,\lr\mm}|^2.
\label{eq:rorbchange}
\ee
Note that when $\omi<0$, the mode energy can be either positive or
negative ($\Delta E_{\alpha}\propto \omi\varepsilon_\alpha$),
and thus $(\Delta N_{orb})_{\alpha,lm}$ can have both signs.
This implies that the resonance can either slow down or speed up the
inspiral. 

The dominant quadrupolar gravitational wave emitted by the binary 
has frequency $\nu_{\rm gw}=2\nu_{\rm orb}=\Omo/\pi$. Thus at 
the $(\alpha,m)$-resonance, the gravitational wave frequency is 
\be
\nu_{\rm gw}={2\over m}|\nui|,
\ee
where $\nui=\omi/(2\pi)$ is the mode frequency in the inertial
frame. In the following
(\S 3), we shall express our results in terms of $\nu_{\rm gw}$ instead 
of $|\nui|$. 

The phase error in the gravitational wave due to the resonance is
$(\Delta\Phi_{\rm gw})_{\alpha,m}=4\pi(\Delta N_{\rm orb})_{\alpha,m}$.
We expect that, for $(\Delta\Phi_{\rm gw})_{\alpha,m}\sim 1$,
the phase error will not affect gravitational wave detection
but could affect detailed analyses of the signals, such as
parameter extraction (E. Flanagan 1998;
S. Hughes 1999, private communication).

The orbital change, or phase error, does not occur 
instantaneously at $D = D_\alpha$.  The duration of the resonance is 
given by
\begin{displaymath}
(\delta t)_{\alpha,m}
\simeq\int\!dt\, e^{im\Phi-i|\omi| t}
\simeq\int\!dt\,e^{-\frac{1}{2}i\mr\dot\Omo t^2}
\end{displaymath}
\be
\, \quad \qquad \simeq\lp\!\frac{2\pi}{\mr\dot\Omo}\!\rp_{\!\alpha}^{\!1/2}
= \lp \frac{2}{3\mr} t_D t_{\rm orb}\rp_{\!\alpha}^{\!1/2}.
\ee
This yields the number of orbital cycles during which resonance
occurs,
\begin{displaymath}
(\delta N_{\rm orb})_{\alpha,m} \simeq \lp\frac{\Omo}{2\pi}\rp_{\!\!\alpha}
(\delta t)_{\alpha,m}
\end{displaymath}
\be
\quad \qquad \qquad = \, 15\, M_{1.4}^{-5/6}
\lp\!\frac{1+q}{q^3}\!\rp^{\!\!1/6}\!\!\!
\mr^{-1/2}\!\lp\!\!\frac{\nu_{\rm gw}}{100\mbox{ Hz}}\!\!\rp^{\!-5/6}
\!\!\!\!\!\!\!\!,
\ee
where $M_{1.4}$ is the NS mass in units of $1.4M_\odot$
and recalling that $\nu_{\rm gw} = (2/\mr)\nui$.
The resonance can be considered instantaneous if 
$(\delta N_{\rm orb})_{\alpha,m}$ is much less than
$dN_{\rm orb}/d\ln\nu_{\rm gw}$, the characteristic number of
orbital cycles the binary spends near orbital frequency
$\nu_{\rm orb}=\nu_{\rm gw}/2$, as given by 
\begin{eqnarray}
{dN_{\rm orb}\over d\ln \nu_{\rm gw}} & = &
\frac{5c^5}{192\pi G^{5/3}}\frac{1}{M^{5/3}}\lp\frac{1+q}{q^3}\rp^{1/3}
\!\!\!\!\frac{1}{(\pi\nu_{\rm gw})^{5/3}} \nonumber \\
& = & 229 \, M_{1.4}^{-5/3} \lp\frac{1+q}{q^3}\rp^{1/3}
\lp\frac{\nu_{\rm gw}}{100 \mbox{Hz}}\rp^{-5/3}\!\!\!\!.
\end{eqnarray}

\setcounter{equation}{0}

\section{Aligned or Anti-Aligned Spin-Orbit ($\beta = 0^\circ$ or 
$180^\circ$)}

We now apply the general equations derived in
\S 2 to calculate the effects of resonance for different types of modes 
in coalescing binary neutron stars. In this section we consider the cases
when ${\bf\Omega}_s$ and ${\bf\Omega}_{\rm orb}$ are either aligned
($\beta=0^\circ$) or anti-aligned ($\beta=180^\circ$). 
Various mode properties ($\omi,~Q_{\alpha,lk}$, and $\varepsilon_\alpha$)
and our procedure to calculate them are summarized. The case
of general $\beta$ will be discussed in \S 4.

For $\beta = 0^\circ$ or $180^\circ$, we clearly have $m=k$, i.e., 
a given mode $\alpha=\{\lm\mm\}$ can only be excited once, at the
orbital frequency given by $k\Omo=|\omi|$. 

\subsection{F-modes}

For nonrotating NSs, the f-modes have frequencies of
order $(GM/R^3)^{1/2}$, which are too large for 
any resonance to occur during binary inspiral. Rotation can 
reduce the mode frequencies, making resonant excitations possible. 
In this paper, we shall consider only the f-modes with $j=k$.
Since $\omi = \omr - k \Oms$, these $j=k$ modes are more strongly affected
by rotation. Only the modes with $\omr>0$ need to be considered, since
rotation actually increases $|\omi|$ for the $\omr<0$ modes,
making their resonances impossible (see \S 2.2.1).
 
To calculate the f-mode properties relevant for tidal resonance, 
we shall model the NS as an incompressible Maclaurin spheroid,
whose equatorial radius and polar radius are $a_1$ and $a_3$,
respectively. The spin frequency $\Oms$ is related to the eccentricity 
$e=(1-a_3^2/a_1^2)^{1/2}$ of the spheroid via the expression
\begin{displaymath}
\Oms^2 = \frac{2\pi G \rho}{e^3}
\end{displaymath}
\be
\qquad \times \lb \lp
1-e^2 \rp^{1/2}\lp 3-2e^2 \rp \sin^{-1}e - 3e \lp 1-e^2 \rp \rb,
\label{eq:omegasf}
\ee
where $\rho=3M/(4\pi R^3)$ is the (uniform) density of the spheroid,
and $R=(a_1^2a_3)^{1/3}$ is the mean radius. 
The frequency of the $j=k$ mode in the rotating frame has been
obtained by Bryan (1889) and Comins (1979):
\be
\frac{\omr}{\Oms} = 1 \pm \left\{1 - \frac{2\mm e^2 R_\mm}
{\lb \lp 3-2e^2 \rp \sin^{-1}e - 3e \lp 1-e^2 \rp^{1/2} \rb}
\right\}^{\!\!1/2}\!\!\!\!\!\!\!, \label{eq:omegarf}
\ee
where the $\pm$ sign refers to modes which propagate opposite to
$\Oms$ and in the same direction as $\Oms$, respectively, and
\begin{eqnarray}
R_\mm & = & \frac{(1-e^2)^{1/2}}{e}\frac{(2\mm-1)!!}{(2\mm)!!}
\!\!\!\sum^{\infty}_{p=\mm+1}
\!\!\frac{(2p-2)!!}{(2p-1)!!}e^{2(p-\mm)} \nonumber \\
&& +\frac{(1-e^2)}{e^2}\lb \sin^{-1}e
- \frac{e}{(1-e^2)^{1/2}}\rb.
\end{eqnarray}
For $\lm=\mm=2$, a more compact expression for the mode
frequency is given by Chandrasekhar (1969) as
\be
\omr=\Oms \pm \sqrt{4\pi G\rho B_{11} - \Oms^2},
\ee
where $B_{11}$ is the index symbol (a dimensionless function of
$e$) as defined in Chandrasekhar (1969).
Also note that for $\Oms=0$, the mode frequency becomes
\be
\omr = \pm \sqrt{\frac{2GM}{R^3}\frac{\lm(\lm-1)}{2\lm+1}}.
\ee
This is essentially the Kelvin mode.
As discussed above, we shall only consider the mode with $+$
sign in (\ref{eq:omegarf}).

\begin{figure*}
\hbox{\epsffile{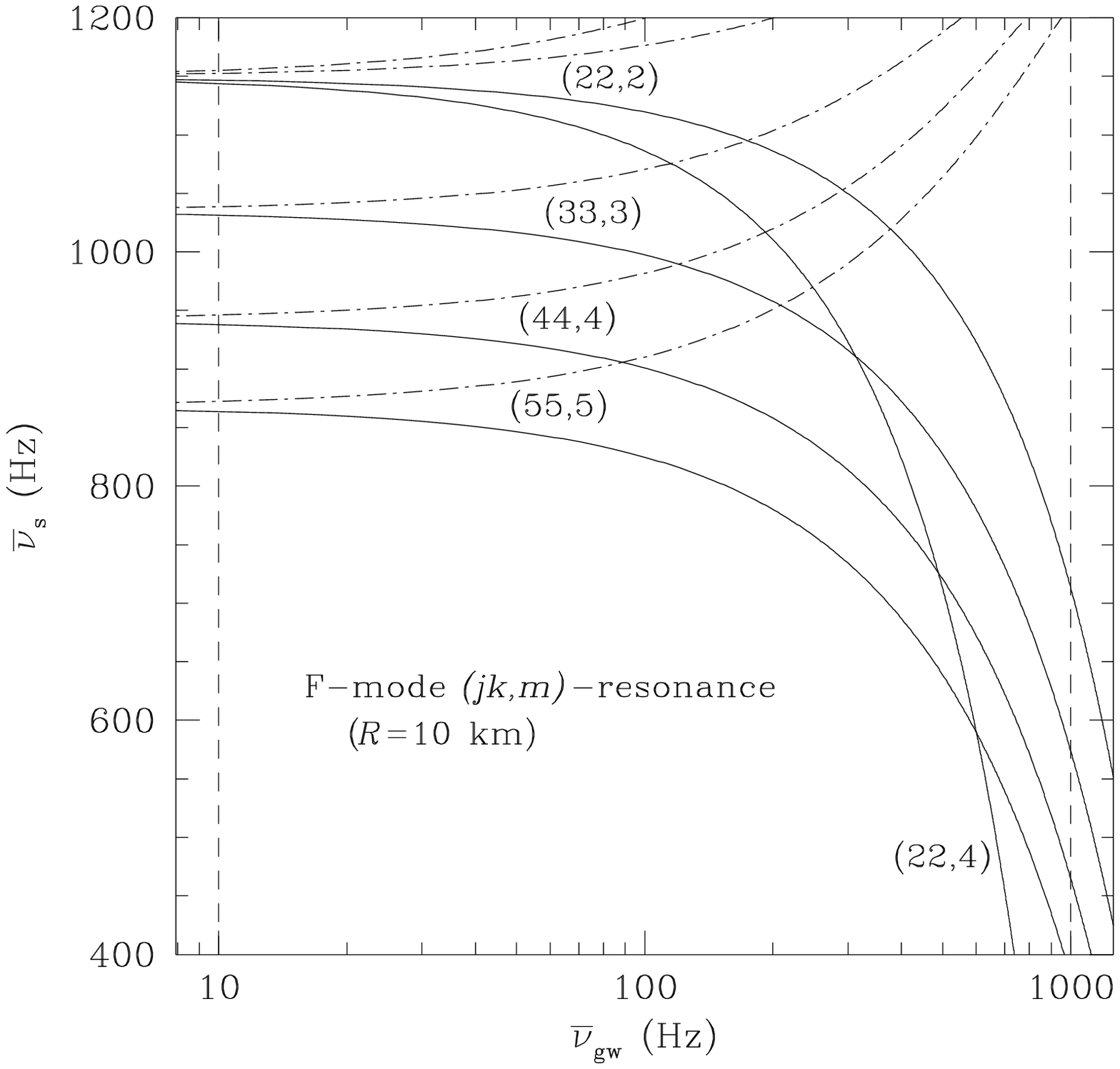}}
\caption{Normalized spin frequency, ${\bar \nu_s} = \nu_s\,M_{1.4}^{-1/2}
R_{10}^{3/2}$, of the NS as a function of the normalized gravitational
wave frequency, ${\bar \nu}_{\rm gw}$, for the f-mode
$(\lm\mm,\mr)$-resonance (taking $\lm = \mm =$ 2--5), where
${\bar \nu_{\rm gw}}$ is related to the f-mode frequency by
${\bar \nu_{\rm gw}} = \nu_{\rm gw}\,M_{1.4}^{-1/2} R_{10}^{3/2}
= (2/\mr)|\nui|\,M_{1.4}^{-1/2} R_{10}^{3/2}$.  The labels in the
figure give the values of $(\lm\mm,\mr)$.  The solid lines are for
the $\omr > 0$ and $\omi > 0$ (spin-retrograde) stable modes.
The dot-dashed lines are for the $\omr > 0$ and $\omi < 0$
(spin-prograde) unstable modes.  Note that the (22,4)-resonance is
possible only for misaligned spin-orbit configurations
($\beta\ne 0^\circ,180^\circ$).  It is also clear the unstable modes
occur at extremely high spin rates.  The vertical dashed lines
indicate the frequency range of gravitational wave detectors,
$\nu_{\rm gw} =$ 10--1000~Hz.} \label{fig:fspin}
\end{figure*}

Figure~\ref{fig:fspin} shows the normalized spin frequency,
\be
\bar\nu_s=\nu_s M_{1.4}^{-1/2}R_{10}^{3/2} = 2170 \, {\hat\Omega_s}~{\rm Hz},
\ee
where $R_{10}$ is the NS radius in units of $10$~km and
$\hat \Oms = \Oms\lp R^3/GM\rp^{1/2}$,
plotted against the normalized gravitational wave frequency,
\be
\bar\nu_{\rm gw}=\nu_{\rm gw}M_{1.4}^{-1/2}R_{10}^{3/2}
=2170\,\left({2\over m}\right)|\hat\omi|~{\rm Hz},
\ee
at which resonance occurs [see eq.~(\ref{con1})] for several different 
f-modes ($j=k=m=2,3,4,5$).
Higher order modes ($k>5$)
are not considered since, as shown below, their contribution
to the orbital change is negligible. 
The spin frequency at which the star becomes secularly unstable
(in the absence of viscosity) is where
$\omi$ becomes negative; for larger $\mm$-values, secular instability 
occurs at lower $\Oms$. Resonance can also occur at these rotation 
frequencies. For completeness, we shall include these unstable modes,
even though they would require very high spin rates.

The two vertical dashed lines in Fig.~\ref{fig:fspin}
indicate the frequency range of gravitational wave detectors,
$\nu_{\rm gw} = 10-1000$~Hz.
Note that the binary merger occurs at about $r_{\rm min}=3R$
(for equal mass NSs), corresponding to the maximum gravitational
wave frequency
\be
\nu_{\rm gw}^{(\rm max)} = 1181\,M_{1.4}^{1/2}R_{10}^{-3/2}
\lp\frac{3R}{r_{\rm min}}\rp^{3/2}~{\rm Hz}.
\label{eq:mergefreq}
\ee
Near this frequency, the distortion to the NS(s) is significant.
Clearly for resonance to occur before merger would require
$\nu_{\rm gw} < \nu_{\rm gw}^{(\rm max)}$.

To calculate the tidal coupling coefficient $Q_{\alpha,lk}$, we
need to determine the Lagrangian displacement, $\bxi_\alpha$. 
We begin with the Euler equation in the corotating frame for
the inviscid fluid:
\be
\frac{\partial \bv}{\partial t} + \lp \bv \cdot \nabla \rp \bv
+ 2\bOmega_s \times \bv = - \nabla \Psi,
\label{euler}\ee
where
\be
\Psi = \frac{p}{\rho} + \Phi - \frac{1}{2} \Oms^{2} (x^2 + y^2).
\ee
$\Phi$ is the self-gravitational potential.
Linearizing and assuming small perturbations $\bv,\Psi\propto e^{i\omr t}$,
eq.~(\ref{euler}) reduces to
\be
i \omr \bv + 2 \bOmega_s \times \bv = - \nabla \Psi. \label{eq:euler}
\ee
Combining this with the continuity equation for an incompressible fluid,
$\nabla\cdot\bv=0$, we find
\be
\frac{\partial^2 \Psi}{\partial x^2} + \frac{\partial^2 \Psi}{\partial y^2}
+ \left\{ 1 - \frac{4 \Oms^2}{[\omr]^2} \right\} \frac{\partial^2 \Psi}{\partial z^2} 
= 0.
\ee
A specific set of solutions to this equation, corresponding to the 
$j=k$ mode, is
\be
\Psi = A_\mm (x + i y)^{\mm-1}e^{i\omr t}, \label{eq:phisoln}
\ee
where $A_\mm$ is a normalization constant.
The Lagrangian displacement $\bxi=\bv/(i\omr)$ can be obtained by 
substituting (\ref{eq:phisoln}) into (\ref{eq:euler}), giving
\be
\bxi = {kA_k\over \omr(\omr-2\Oms)}
\lp x + i y \rp^{\mm-1} \lp \hat{x} + i \hat{y} \rp e^{i\omr t}.
\ee
Adopting the normalization condition (\ref{eq:xinorm}), we can write 
the displacement vector $\bxi$ as
\begin{eqnarray}
\bxi(x,y,z) & = & \lb \frac{(2\mm+1)!!}{(2\mm)!!}
\frac{1}{4\pi \rho \mm a_1^{2\mm} a_3}\rb^{1/2} \nonumber \\
&& \times \lp x + i y \rp^{\mm-1} \lp \hat{x} + i \hat{y} \rp e^{i\omr t},
\end{eqnarray}
\begin{eqnarray}
\bxi(r_\perp,\phi,z) & = & \lb \frac{(2\mm+1)!!}{(2\mm)!!}
\frac{1}{4\pi \rho \mm a_1^{2\mm} a_3}\rb^{1/2} \nonumber \\
&& \times r_\perp^{\mm-1} e^{i(\mm \phi + \omr t)}
\lp \hat{r}_\perp + i \hat{\phi} \rp,
\end{eqnarray}
\be
\bxi(r,\theta,\phi) = \lp \xi_r, \xi_h \frac{\partial}{\partial \theta},
\xi_h \frac{1}{\sin \theta} \frac{\partial}{\partial \phi} \rp
Y_{\mm \mm}(\theta,\phi) e^{i\omr t},
\label{spherical}\ee
in Cartesian, cylindrical, and spherical coordinates, respectively. 
In (\ref{spherical}), the radial and horizontal wavefunctions are
given by [neglecting the factor $(-1)^k$]
\be
\xi_r = \lp \frac{\mm}{\rho a_1^{2\mm} a_3} \rp^{1/2} r^{\mm-1},
\quad \xi_h = \frac{\xi_r}{\mm},
\ee
and the spherical harmonic function $Y_{kk}$ is given by 
\be
Y_{kk}(\theta,\phi)={(-1)^k\over 2^k k!}\sqrt{(2k+1)!\over 4\pi}
(\sin\theta)^k e^{ik\phi}.
\ee
Applying appropriate boundary conditions (Comins 1979)
for the eigenfunction then yields the expression for the mode frequency 
$\omr$ as given by (\ref{eq:omegarf}).

Using the spherical representation of $\bxi$ for the f-modes,
the tidal coupling coefficient (for $\alpha=\{kk\}$ and
$l=m'=k$), defined by (\ref{eq:qcoup}), can be written as 
\begin{displaymath}
Q_{kk,kk}=\lp \frac{\mm}{\rho a_1^{2\mm} a_3} \rp^{1/2}\int\!d^3x\,\rho
r^{2k-2}
\end{displaymath}
\be
\qquad \times \left[k\left|Y_{kk}\right|^2
+{1\over k}\left|{\partial Y_{kk}\over\partial\theta}
\right|^2+{1\over k\sin^2\theta}\left|{\partial Y_{kk}\over\partial\phi}
\right|^2\right].
\label{qexpression}
\ee
Note that for a given $\theta$, the radial variable ranges from
$0$ to $r_{\rm max}(\theta)$, with
\be
r_{\rm max}(\theta) = \lp \frac{\sin^2\theta}{a_1^2}
+ \frac{\cos^2\theta}{a_3^2} \rp^{-1/2} \!\!\!\!\!\!
 = {a_1}\lp 1 + \frac{\mu'^2}{\zeta^2} \rp^{-1/2}\!\!\!\!\!\!,
\ee
where $\mu' = \cos \theta$ and $\zeta = \lp 1-e^2 \rp^{1/2}/e$.
Equation (\ref{qexpression}) can then be evaluated explicitly, giving
\be
Q_{kk,kk} = \lp\frac{3\mm}{4\pi}\rp^{1/2}\lp\frac{a_1}{R}\rp^{\mm-1},
\label{eq:qcoupf}
\ee
recalling the normalization of $Q_{\alpha,lk}$ in units such that $G=M=R=1$.
Furthermore,
\be
\hat{T}_\alpha = -(\mm-1)\hat \Oms
\ee
\be
\hat{\varepsilon}_\alpha = \hat \omi + (\mm-1)\hat \Oms
=\hat\omr-\hat\Oms.
\label{eq:epsilf}
\ee

\begin{figure*}
\hbox{\epsffile{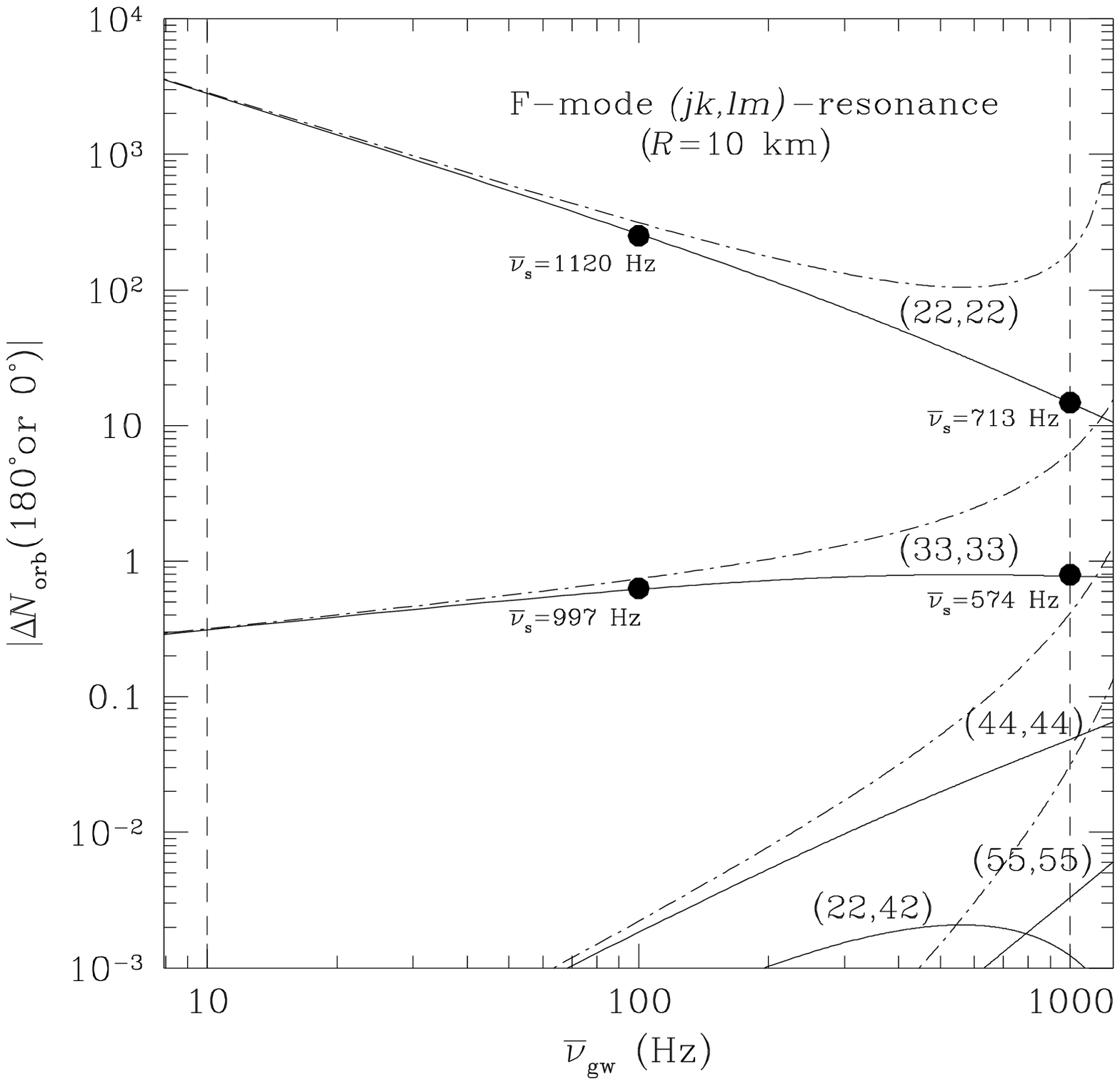}}
\caption{Orbital cycle change $|(\Delta N_{\rm orb})_{\lm\mm,\lr\mr}|$,
with $M=1.4 M_\odot$, $R=10$ km, $q = 1$, as a function of the
normalized gravitational wave frequency, ${\bar \nu_{\rm gw}}
= \nu_{\rm gw}\,M_{1.4}^{-1/2} R_{10}^{3/2}
= (2/\mr)|\nui|\, M_{1.4}^{-1/2} R_{10}^{3/2}$,
for the f-mode $(\lm\mm,\lr\mr)$-resonance.  The labels in the figure
give the values of $(\lm\mm,\lr\mr)$.  The solid lines are for the
$\omr > 0$ and $\omi > 0$ (spin-retrograde) stable modes with
$(\Delta N_{\rm orb})_{\lm\mm,\lr\mr}(180^\circ) < 0$.  The dot-dashed
lines are for the $\omr > 0$ and $\omi < 0$ (spin-prograde) unstable
modes with $(\Delta N_{\rm orb})_{\lm\mm,\lr\mr}(0^\circ) > 0$.  Note
that both the (22,22) and (22,42) curves belong to the same
(22,2)-resonance.  The (22,42) curve serves as a baseline for the
(22,44)-resonance (see \S 4 and Fig.~7).  Also recall from Fig.~1
that the unstable modes require extremely high spin rates.
The large dots with labels indicate the spin rate at that gravitational
wave frequency.
The vertical dashed lines indicate the frequency range of gravitational
wave detectors, $\nu_{\rm gw} =$ 10--1000~Hz.} \label{fig:forbchange}
\end{figure*}

Substituting equations (\ref{eq:omegarf}), (\ref{eq:qcoupf}), and
(\ref{eq:epsilf}) into (\ref{eq:orbchange}) or (\ref{eq:rorbchange}),
the change in the number of orbital cycles
due to the resonance can be readily calculated. Figure \ref{fig:forbchange}
depicts $(\Delta N_{\rm orb})_{kk,kk}$ as a function of 
the normalized gravitational wave frequency ${\bar\nu}_{\rm gw}$.
For $\mr = \mm = 2$, we have
\begin{displaymath}
(\Delta N_{\rm orb})_{22,22}
\lp\begin{array}{c}\!\!\!\!180^\circ\!\!\!\! \\ \!\!0^\circ\!\! \end{array}\rp
\simeq \mp\,234\,M_{1.4}^{-4.5} R_{10}^{3.5} \frac{1}{q(1+q)}
\end{displaymath}
\be
\qquad \qquad \qquad \qquad \times
\lp\frac{0.895}{{\hat \varepsilon}_\alpha}\rp
\left|\frac{Q_{22,22}}{0.69}\right|^2
\lp\frac{100\mbox{Hz}}{\nu_{\rm gw}}\rp,
\label{eq:forb2222}
\ee
where we have normalized ${\hat \varepsilon}_\alpha$ and $Q_{22,22}$ to
their respective nonrotating values (rotation can change these
by a factor of $\sim 20$ and $\sim 1.5$, respectively, for the
range of ${\hat\Oms}\lo 0.57$ considered in this paper).
The upper sign (negative $\Delta N_{\rm orb}$) in (\ref{eq:forb2222})
applies when $\omi>0$ and $\beta=180^\circ$, while the lower sign (positive
$\Delta N_{\rm orb}$) applies when $\omi<0$ and $\beta=0^\circ$.
Note that $(\Delta N_{\rm orb})_{kk,kk}$ diverges for $k=2$ as $\nu_{\rm gw}$
decreases. The reason for this is that at larger orbital radii (small
resonant frequencies), the orbit decays more slowly,
and the binary spends a longer time in resonance, giving rise to 
larger $\Delta N_{\rm orb}$. For higher $k$-resonances, the same effect is 
also operative, but the tidal potential is weaker so that 
the net effect is that $(\Delta N_{\rm orb})_{kk,kk}
\propto \nu_{\rm gw}^{(4k-11)/3}$ decreases with decreasing $\nu_{\rm gw}$. 

It is evident from Fig.~\ref{fig:forbchange} that the
$(\lm\mm,\mr) = (22,2)$ and $(33,3)$-resonances contribute significantly to 
the orbital phase evolution. But, to attain such resonances during 
the binary inspiral, the NS of canonical mass ($1.4M_\odot$) and
radius ($10$~km) must be rotating very rapidly, i.e.,
$\nu_s\go 710$~Hz for the $(22,2)$-resonance,
$\nu_s\go 570$~Hz for the $(33,3)$-resonance,
$\nu_s\go 460$~Hz for the $(44,4)$-resonance, and
$\nu_s\go 380$~Hz for the $(55,5)$-resonance
(see Fig.~\ref{fig:fspin}). Such rapid rotation
is physically possible but may be astrophysically unlikely.
Observations of NS binary systems and the current theories
regarding the formation of these systems indicate that such rapid
rotation of the NS is not possible (see \S 5).
If such resonances indeed occur, then their effect on the orbital phase
of the binary inspiral must be taken into account when constructing 
theoretical wave templates in the search for gravitational waves from 
these systems.

\begin{figure*}
\hbox{\epsffile{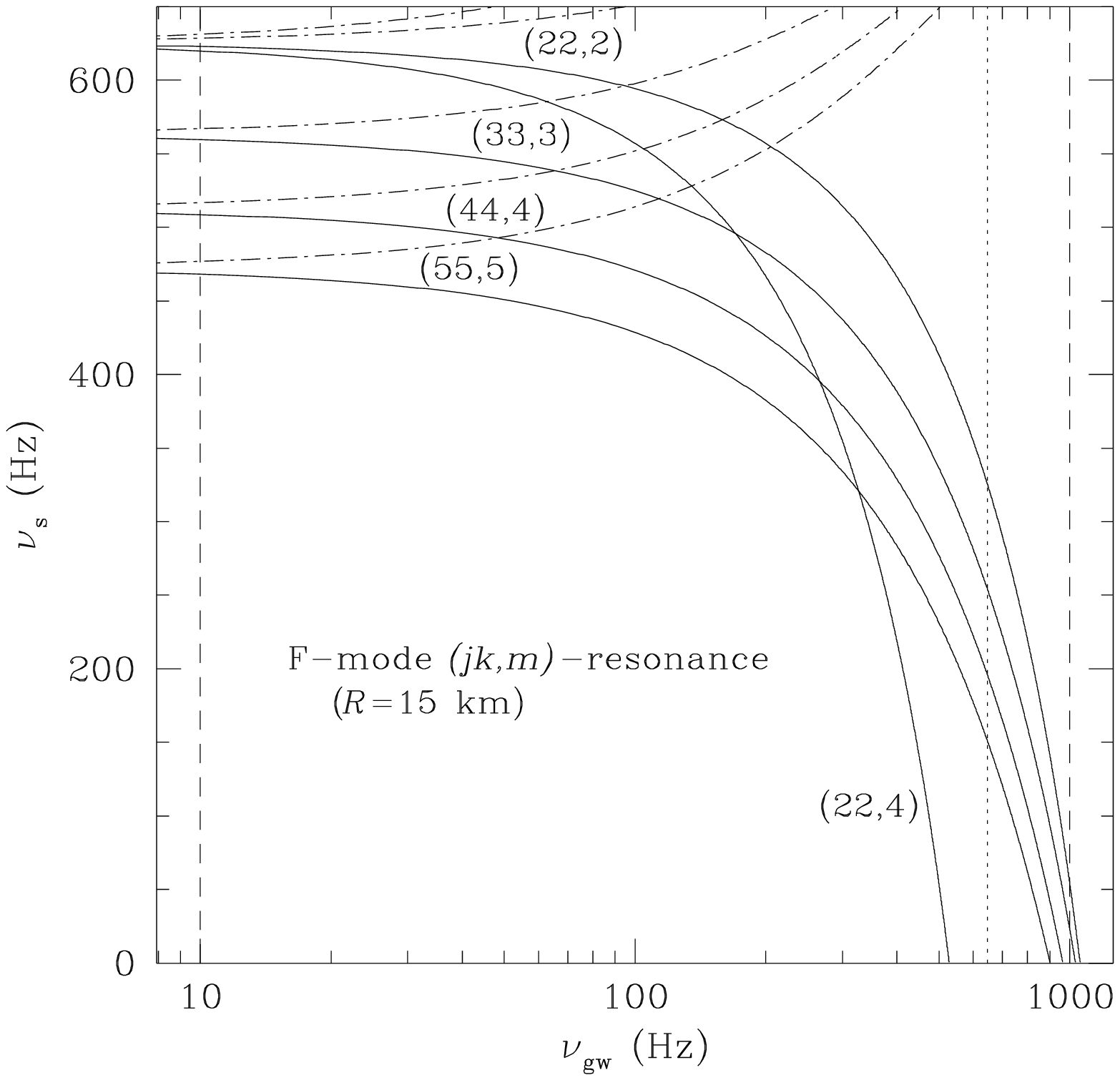}}
\caption{Spin frequency, $\nu_s$, of the $R=15$ km NS
as a function of the gravitational
wave frequency, $\nu_{\rm gw}$, for the f-mode
$(\lm\mm,\mr)$-resonance (taking $\lm = \mm =$ 2--5), where
$\nu_{\rm gw}$ is related to the f-mode frequency by
$\nu_{\rm gw} = (2/\mr)|\nui|$.  The labels in the
figure give the values of $(\lm\mm,\mr)$.  The solid lines are for
the $\omr > 0$ and $\omi > 0$ (spin-retrograde) stable modes.
The dot-dashed lines are for the $\omr > 0$ and $\omi < 0$
(spin-prograde) unstable modes.  Note that the (22,4)-resonance is
possible only for misaligned spin-orbit configurations
($\beta\ne 0^\circ,180^\circ$).  It is also clear the unstable modes
occur at extremely high spin rates.
The vertical dotted line at $\nu_{\rm gw} =$ 640~Hz denotes the
maximum gravitational wave frequency allowed by eq.~(3.8),
which sets a lower limit to the spin frequency.
The vertical dashed lines
indicate the frequency range of gravitational wave detectors,
$\nu_{\rm gw} =$ 10--1000~Hz.} \label{fig:fspin15}
\end{figure*}

\begin{figure*}
\hbox{\epsffile{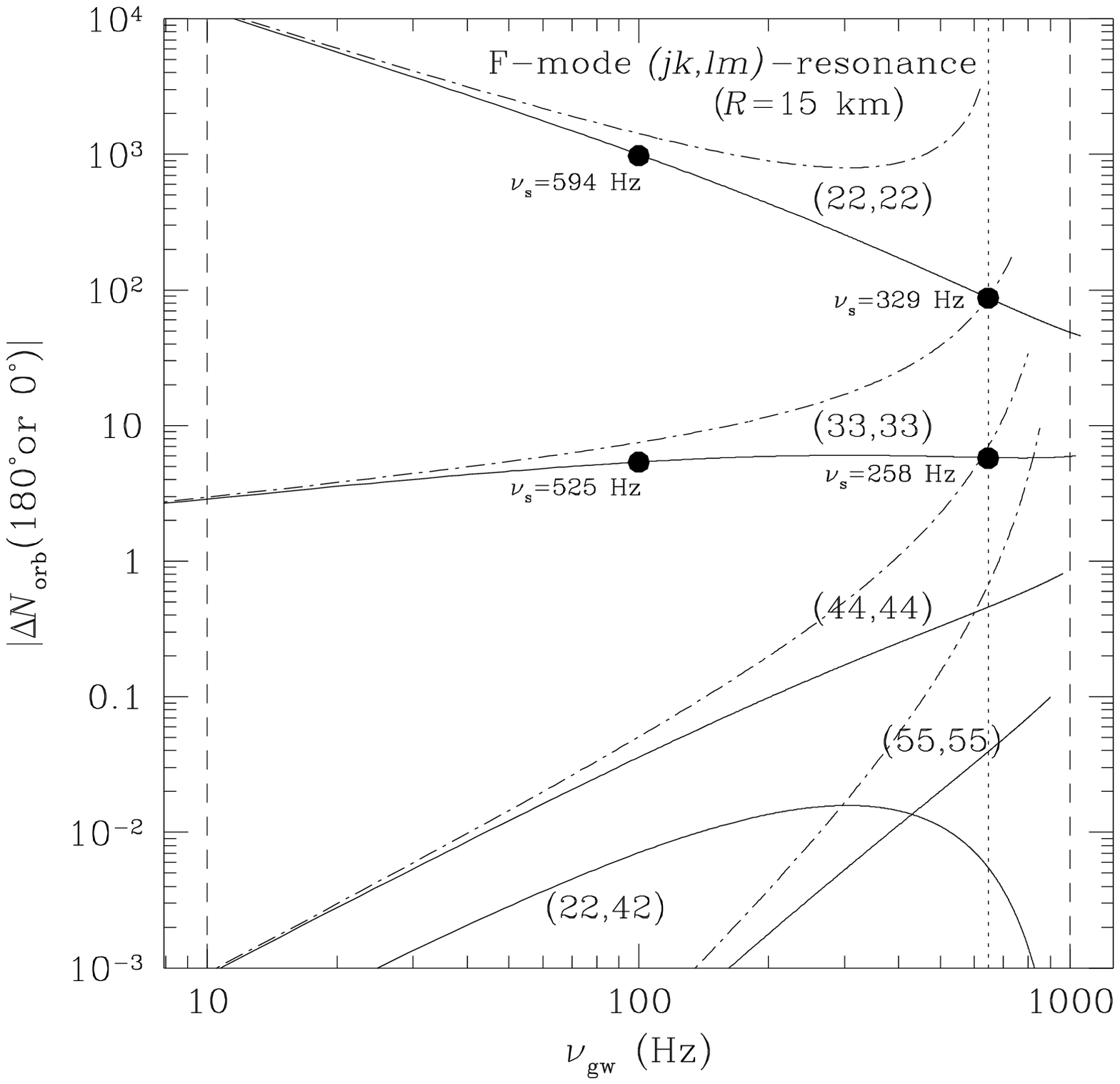}}
\caption{Orbital cycle change $|(\Delta N_{\rm orb})_{\lm\mm,\lr\mr}|$,
with $M=1.4 M_\odot$, $R=15$ km, $q = 1$, as a function of the
gravitational wave frequency, $\nu_{\rm gw} = (2/\mr)|\nui|$,
for the f-mode $(\lm\mm,\lr\mr)$-resonance.  The labels in the figure
give the values of $(\lm\mm,\lr\mr)$.  The solid lines are for the
$\omr > 0$ and $\omi > 0$ (spin-retrograde) stable modes with
$(\Delta N_{\rm orb})_{\lm\mm,\lr\mr}(180^\circ) < 0$.  The dot-dashed
lines are for the $\omr > 0$ and $\omi < 0$ (spin-prograde) unstable
modes with $(\Delta N_{\rm orb})_{\lm\mm,\lr\mr}(0^\circ) > 0$.  Note
that both the (22,22) and (22,42) curves belong to the same
(22,2)-resonance.  The (22,42) curve serves as a baseline for the
(22,44)-resonance (see \S 4 and Fig.~7).  Also recall from Fig.~2
that the unstable modes require extremely high spin rates.
The vertical dotted line at $\nu_{\rm gw} =$ 640~Hz denotes the
maximum gravitational wave frequency allowed by eq.~(3.8).
The large dots with labels indicate the spin rate at that gravitational
wave frequency.
The vertical dashed lines indicate the frequency range of gravitational
wave detectors, $\nu_{\rm gw} =$ 10--1000~Hz.} \label{fig:forbchange15}
\end{figure*}

It can also be seen from eqs.~(\ref{eq:omegasf}) and (\ref{eq:omegarf})
that the NS spin frequency and mode frequencies depend on the radius of the
NS and thus the NS equation of state (EOS).  Current equations of state
allow NSs with $R=15$~km (e.g., Shapiro \& Teukolsky 1983).
This implies that the spin and mode frequencies are lowered, and
resonance occurs at these lower frequencies
as illustrated by Figure~\ref{fig:fspin15}.
Figure~\ref{fig:forbchange15} shows the corresponding
$(\Delta N_{\rm orb})_{kk,kk}$ as a function of the $\nu_{\rm gw}$
for the $R=15$ km NS.
The lower limit to the spin frequency required for resonance
is set by the associated gravitational wave
frequency given by eq.~(\ref{eq:mergefreq}).  This limit is indicated
by the dotted lines in Fig.~\ref{fig:fspin15} and Fig.~\ref{fig:forbchange15}
and yields
$\nu_s\go 330$~Hz for the $(22,2)$-resonance,
$\nu_s\go 260$~Hz for the $(33,3)$-resonance,
$\nu_s\go 200$~Hz for the $(44,4)$-resonance, and
$\nu_s\go 150$~Hz for the $(55,5)$-resonance.

It is necessary to check that the higher-$l$ contribution to the
$(kk,m=k)$-resonance is negligible. Except for the $\Omega_s=0$ case,
the $l=k+2,k+4,\cdots$ tides can also contribute to 
$(\Delta N_{\rm orb})_{kk,k}$. The mode frequency is the same,
but the new tidal coupling coefficient must be calculated.
We shall consider just the (22,42)-resonance and the
(33,53)-resonance here since the higher order terms are much smaller.
Using the mode eigenfunction (\ref{spherical}), we find
\begin{eqnarray}
Q_{22,42} & = & \frac{45}{112}\!\lb\frac{9}{2\pi\lp1-e^2\rp}\rb^{1/2}
\lp\frac{a_1}{R}\rp^3 \nonumber \\
&& \times \int^{1}_{-1}\frac{-5\mu'^4+6\mu'^2-1}
{\lp 1 + \frac{\mu'^2}{\zeta^2} \rp^{7/2}}\,d\mu' \label{eq:fq2242}
\end{eqnarray}
and
\begin{eqnarray}
Q_{33,53} & = & \frac{35}{96}\!\lb\frac{11}{\pi\lp1-e^2\rp}\rb^{1/2}
\lp\frac{a_1}{R}\rp^4 \nonumber \\
&& \times \int^{1}_{-1}\frac{7\mu'^6-15\mu'^4+9\mu'^2-1}
{\lp 1 + \frac{\mu'^2}{\zeta^2} \rp^{9/2}}\,d\mu'.
\label{eq:fq3353}
\end{eqnarray}
Both expressions can be evaluated numerically. 
In Fig.~\ref{fig:forbchange} the results for the orbital
change $(\Delta N_{\rm orb})_{22,42}$
is plotted, indicating that it is indeed much smaller than
$(\Delta N_{\rm orb})_{22,22}$\footnote{
Note that to calculate the total orbital change due to the
$(jk,m)=(22,2)$-resonance, one needs to use eq.~(\ref{eq:orbchange1}) which 
includes contributions from all possible $l$-terms. 
But the smallness of $(\Delta N_{\rm orb})_{\rm 22,42}$ implies
$(\Delta N_{\rm orb})_{\rm 22,2}\simeq 
(\Delta N_{\rm orb})_{\rm 22,22}$ to a good approximation.
The quantity $(\Delta N_{\rm orb})_{\rm 22,42}$ will also be used in \S4 as 
a fiducial number for the $(22,44)$-resonance.}.
The orbital change
$(\Delta N_{\rm orb})_{33,53}$ is not shown since it is 
below the limit of interest for this paper.

\subsection{G-modes}

Gravity modes in neutron stars arise from the composition
(proton to neutron ratio)
gradient in the stellar core (Reisenegger \& Goldreich 1992; Lai 1994),
density discontinuities in the crust (Finn 1987), as well as
thermal buoyancy associated with finite temperatures, either due to
internal heat (McDermott et al.~1988) or due to
accretion (Bildsten \& Cutler 1995). In this paper, we shall focus on the core
g-modes since the gradient of proton to neutron ratio in the NS
interior provides the largest buoyancy. For a nonrotating NS,
the typical frequencies of low order g-modes are around $100$~Hz and  
are smaller for higher order modes. Thus we expect that 
even modest stellar rotation rate ($\sim 100$~Hz) can change the 
mode properties dramatically.
 
The core g-modes of rotating NS have been calculated by Lai (1998)
using the traditional approximation. Two NS models
are considered based on different EOS.
Model UU gives a nonrotating NS radius $R=13.47$~km for a $M=1.4M_\odot$,
and its lowest order ($n=1$) $j=2$ g-mode has frequency 
$\nui=148$~Hz and $Q_{22,22}\simeq0.62\times10^{-3}$.
Model AU gives a nonrotating NS radius $R=12.37$~km for a $M=1.4M_\odot$,
and its lowest order ($n=1$) $j=2$ g-mode has frequency 
$\nui=72$~Hz and $Q_{22,22}\simeq0.11\times10^{-3}$.

\begin{figure*}
\hbox{\epsffile{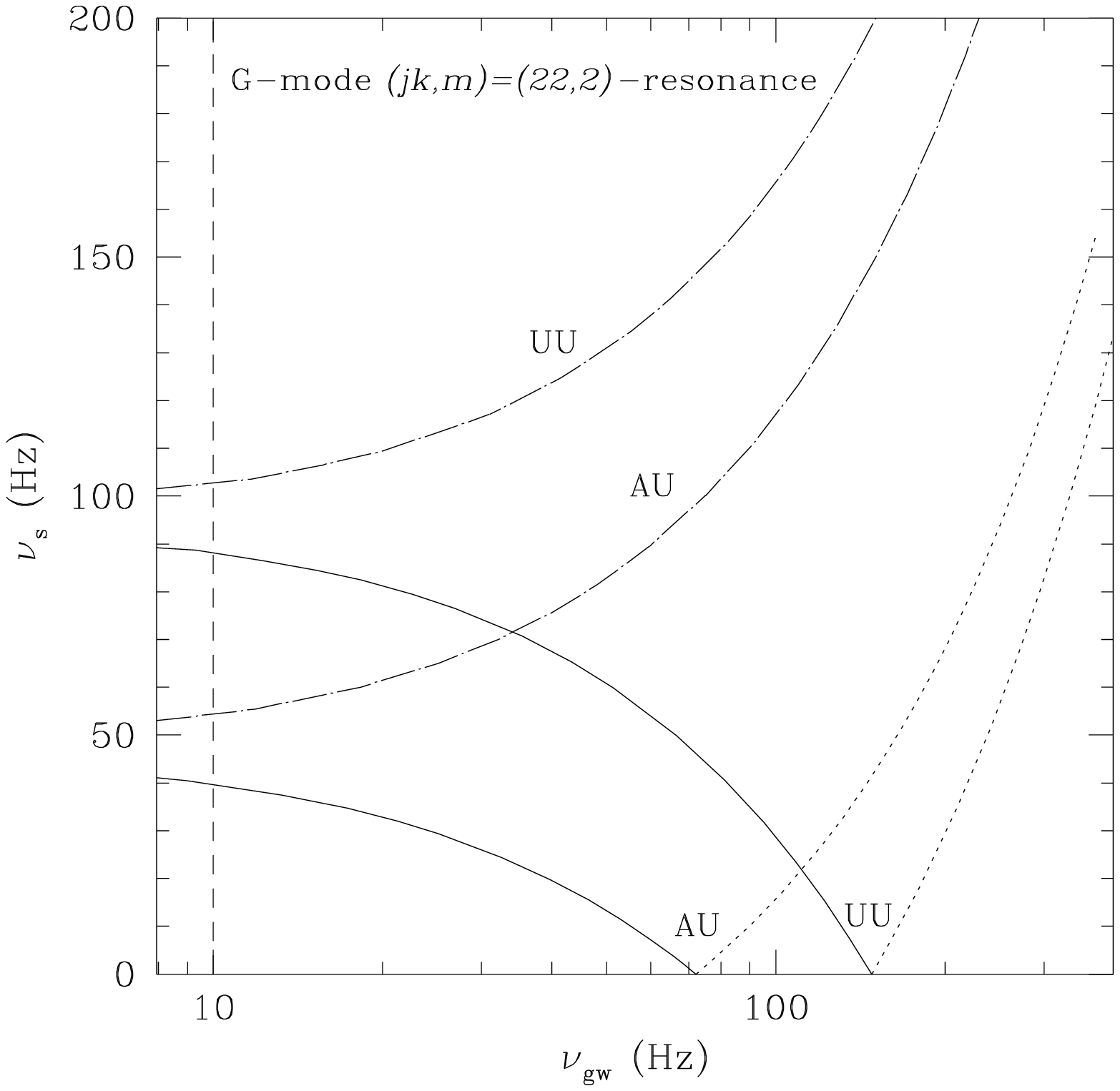}}
\caption{Spin frequency, $\nu_s$, of the NS as a function of
the observed gravitational wave frequency, $\nu_{\rm gw}$, for the
$(\lm\mm,\mr)=(22,2)$-resonance, where $\nu_{\rm gw}$ is related to the
g-mode frequency, $\nui$, by $\nu_{\rm gw} = (2/\mr)|\nui| = |\nui|$.
UU and AU are two models of the NS EOS (see text for the NS parameters)
given in Lai (1998).  The solid lines are for the $\omr > 0$ and $\omi > 0$
(spin-retrograde) stable modes.  The dot-dashed lines are for the
$\omr > 0$ and $\omi < 0$ (spin-prograde) unstable modes.  The dotted
lines are for the $\omr < 0$ and $\omi < 0$ (spin-prograde) stable modes.
The vertical dashed line indicates the lower limit of the frequency
range of gravitational wave detectors, $\nu_{\rm gw} =$ 10--1000~Hz.}
\label{fig:gspin}
\end{figure*}

Figure~\ref{fig:gspin} shows the spin frequency,
$\nu_s$, as a function of the observed gravitational wave frequency,
$\nu_{\rm gw}=(2/m)|\nui|=|\nui|$, for the $m=2$ resonance of
the $\{jk\}=\{22\}$ g-modes ($n=1$).
The $\omr>0$ mode propagates in opposite direction as the spin
[see the discussion following eq.~(\ref{eq:omegar})].  As $\nu_s$
increases, the mode frequency, $\nui$, in the inertial frame decreases; 
$\nui$ can even become negative when $\nu_s$ exceeds a certain
critical value ($95$~Hz for model UU and $47$~Hz for model AU).
This corresponds to the onset of secular instability for
the g-mode (Lai 1998). We include these unstable modes because, as shown
by Lai (1998), viscosity tends to stabilize these modes for
typical parameters of the NS. 

Figure~\ref{fig:gspin} also shows the results for the spin-prograde mode,
which has $\omr<0$ [see the discussion following eq.~(\ref{eq:omegar})].
These are calculated in a similar manner as in Lai (1998). These modes
always have negative $\omi$, but they have positive energy and thus
are secularly stable.  Rotation clearly increases the gravitational
wave frequency at which resonance occurs for these modes. 

\begin{figure*}
\hbox{\epsffile{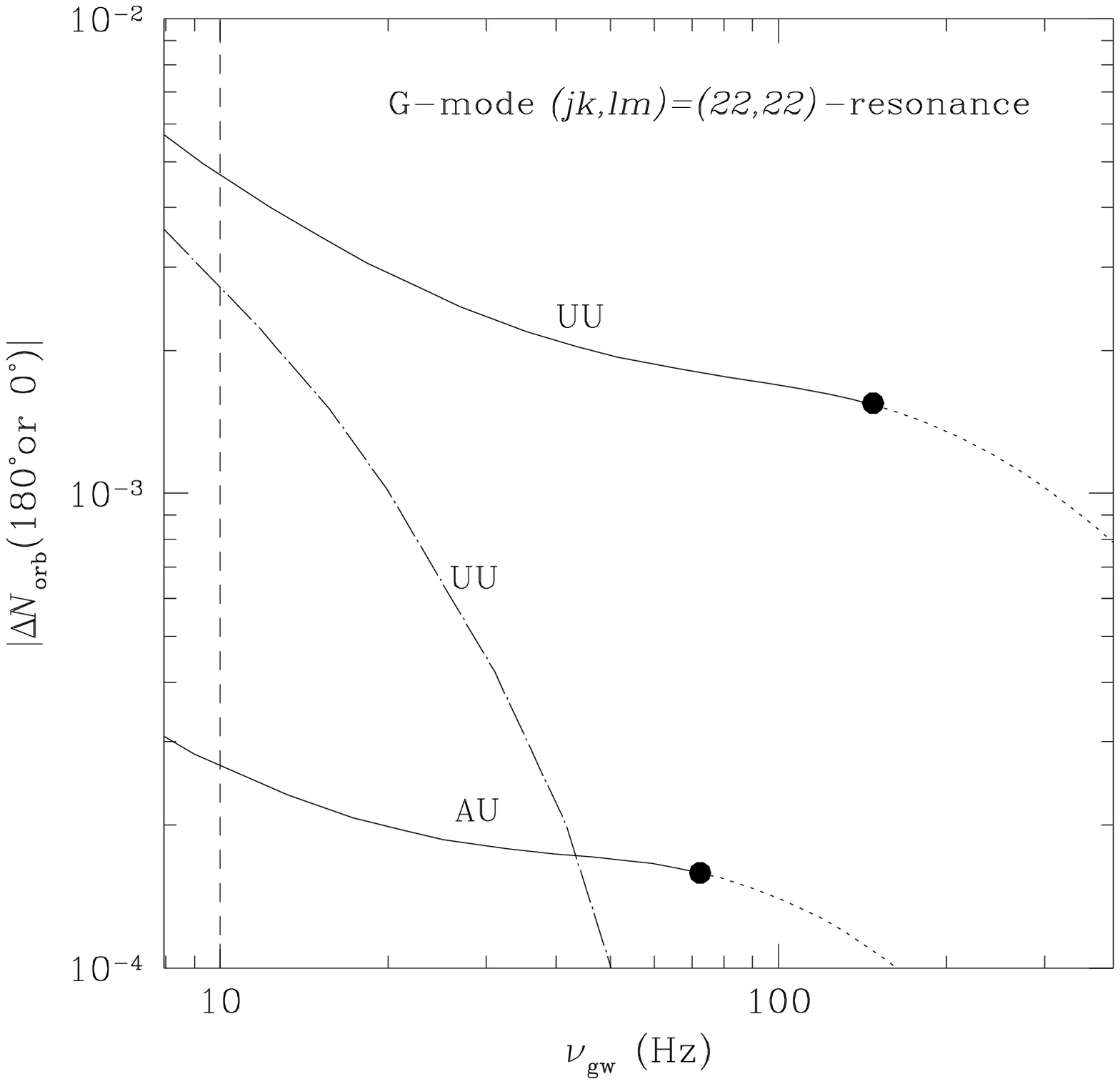}}
\caption{Orbital cycle change $|(\Delta N_{\rm orb})_{22,22}|$, with
$q = 1$, as a function of the observed gravitational wave frequency,
$\nu_{\rm gw} = (2/\mr)|\nui|=|\nui|$, for the g-mode
$(jk,lm)= (22,22)$-resonance.  UU and AU are two models of the NS EOS
(see text for the NS parameters) given in Lai (1998).  The solid lines
are for the $\omr > 0$ and $\omi > 0$ (spin-retrograde) stable modes
with $(\Delta N_{\rm orb})_{22,22}(180^\circ) < 0$.  The dot-dashed
lines are for the $\omr > 0$ and $\omi < 0$ (spin-prograde) unstable
modes with $(\Delta N_{\rm orb})_{22,22}(0^\circ) > 0$ (Note that the
result for model AU is below the limit of the figure).  The dotted
lines are for the $\omr < 0$ and $\omi < 0$ (spin-prograde) stable
modes with $(\Delta N_{\rm orb})_{22,22}(0^\circ) < 0$.  The large
dots indicate where the NS is nonrotating.  The vertical dashed line
indicates the lower limit of the frequency range of gravitational wave
detectors, $\nu_{\rm gw} =$ 10--1000~Hz.} \label{fig:gorbchange}
\end{figure*}

Figure~\ref{fig:gorbchange} shows the orbital change,
$(\Delta N_{\rm orb})_{22,22}$, due to the g-mode
resonance as a function of the observed gravitational wave
frequency, $\nu_{\rm gw}$. The mode parameters $Q_{22,22}$ and 
$\varepsilon_\alpha$ have been obtained using 
the method of Lai (1998). Using eq.~(\ref{eq:orbchange}) or
(\ref{eq:rorbchange}) (depending on the sign of $\omi$), we find
\begin{displaymath}
(\Delta N_{\rm orb})_{22,22}
\lp\begin{array}{c}\!\!\!\!180^\circ\!\!\!\! \\ \!\!0^\circ\!\! \end{array}\rp
\simeq \mp 0.0086\,\frac{1}{q(1+q)}
\end{displaymath}
\be
\quad \qquad \qquad \qquad \times
\lp\frac{0.107}{{\hat \varepsilon}_\alpha}\rp
\left|\frac{Q_{22,22}}{6.2\times10^{-4}}\right|^2
\lp\frac{50 \mbox{Hz}}{\nu_{\rm gw}}\rp
\label{eq:gorb2222uu}
\ee
for model UU and
\begin{displaymath}
(\Delta N_{\rm orb})_{22,22}
\lp\begin{array}{c}\!\!\!\!180^\circ\!\!\!\! \\ \!\!0^\circ\!\! \end{array}\rp
\simeq \mp 0.00045\,\frac{1}{q(1+q)}
\end{displaymath}
\be
\quad \qquad \qquad \qquad \times
\lp\frac{0.046}{{\hat \varepsilon}_\alpha}\rp
\left|\frac{Q_{22,22}}{1.1\times10^{-4}}\right|^2
\lp\frac{50 \mbox{Hz}}{\nu_{\rm gw}}\rp
\label{eq:gorb2222au}
\ee
for model AU.
We have normalized ${\hat \varepsilon}_\alpha$ and $Q_{22,22}$ to
their respective nonrotating values (rotation can change these
by a factor of $\sim2$ for the range of ${\hat\Oms}\lo 0.3$ considered
in this paper).
The upper sign in (\ref{eq:gorb2222uu}) and (\ref{eq:gorb2222au})
applies when $\omi>0$ and $\beta=180^\circ$, while the lower sign
applies when $\omi<0$ and $\beta=0^\circ$.
However, when $\omr<0$ and $\omi<0$, ${\hat \varepsilon}_\alpha<0$;
therefore $\Delta N_{\rm orb}<0$ (a decrease in the number of orbits).

The large dots in Fig.~\ref{fig:gorbchange} correspond
to the case when the NS is nonrotating. 
We see that retrograde rotation ($\beta=180^\circ$) can increase
$|(\Delta N_{\rm orb})_{22,22}|$ for the spin-retrograde mode
by shifting the resonance to smaller orbital frequency, where the
orbital decay is slower.  For the spin-prograde modes, rotation decreases
$|(\Delta N_{\rm orb})_{22,22}|$ from the nonrotating value.
It is clear from Fig.~\ref{fig:gorbchange} that the
orbital change is small, especially in comparison to the orbital
change generated by f-modes.  However, the g-mode resonances
can occur more easily than the f-modes since the
g-mode resonance does not require the rapid rotation of the NS.

The above results apply for particular neutron star models. 
As in eq.~(\ref{eq:forb2222}), we have the scaling:
$(\Delta N_{\rm orb})_{22,22}\propto M_{1.4}^{-4.5}R_{10}^{3.5}$;
thus the orbital change is larger for low-mass NSs which typically have
larger $R/M$.

\subsection{R-modes}

For nonrotating stars, r-modes are ``degenerate'' toroidal modes
with zero frequency and zero density perturbation.  Rotation
breaks the degeneracy via the Coriolis force, giving rise to 
finite-frequency r-modes. For the analysis of the r-modes in 
this section and in \S4.3, we shall consider the slowly-rotating
[$\hat\Omega_s\equiv\Omega_s/(GM/R^3)^{1/2} \ll 1$] 
incompressible NS models. As shown by Provost et
al.~(1981), only the $\lm =\mm$ r-modes exist for barotropic stars.

Following Saio (1982) and Kokkotas \& Stergioulas (1998),
it is convenient to introduce a new radial coordinate $a$ which is
related to the radial distance $r$ through 
$r=a(1+\epsilon)$, where $\epsilon(a,\theta)\propto\hat\Omega_s^2$
represents the centrifugal distortion of the star. In the new coordinates, 
all unperturbed quantities are functions of $a$ only.
The displacement eigenfunction of the $\alpha=\{kk\}$ r-mode can be
written as
\begin{eqnarray}
{\bxi_{kk}\over a}\!\!\!\!&=&\!\!\!\! \left(0,K_{kk}\frac{1}{\sin\theta}
\frac{\partial}{\partial \phi},-K_{kk}
\frac{\partial}{\partial \theta}\right)Y_{kk} \nonumber \\
\!\!\!\!&&\!\!\!\! +\left(S_{k+1 k},H_{k+1 k}{\partial\over\partial\theta},
H_{k+1 k}{1\over\sin\theta}{\partial\over\partial\phi}\right)Y_{k+1 k},
\label{eq:xir}
\end{eqnarray}
where the first term represents the toroidal (axial) component and
the second term represents the spheroidal (polar) component. The 
r-mode is predominantly toroidal, and the
radial eigenfunctions scale with the rotation 
rate as $K_{kk}\sim \hat\Omega_s^0$,
while $S_{k+1 k},H_{k+1 k}\sim \hat\Omega_s^2$. 
For a uniform density star, the mode frequency in the rotating frame
is given, to order $\hat\Omega_s^3$, by
(Saio 1982; Kokkotas \& Stergioulas 1998)
\be
\hat\omr=\frac{2}{\mm+1}\hat\Oms
+ \frac{5\mm(\mm+1)^2-8}{(\mm+1)^4}\hat\Oms^3.
\label{eq:omegarr}
\ee
In the inertial frame, this gives
\be
\hat\omi=-\frac{(\mm+2)(\mm-1)}{\mm+1}\hat\Oms
+ \frac{5\mm(\mm+1)^2-8}{(\mm+1)^4}\hat\Oms^3.
\label{eq:omegair}
\ee
Note that the r-mode always has positive $\omr$ and
negative $\omi$, i.e, the wave pattern propagates
opposite to spin in the rotating frame while it becomes
spin-prograde in the inertial frame, indicating that the r-mode
is secularly unstable, although viscosity tends to 
stabilize the mode for small $\hat\Omega_s$ (Andersson~1998;
Andersson et al.~1998; Friedman \& Morsink 1998; Lindblom et al.~1998).

The eigenfunction (\ref{eq:xir}) implies the density perturbation for
the r-mode $\delta\rho_{kk}\propto Y_{k+1 k}$. Thus the tidal
coupling coefficient, $Q_{kk,lk}$, is nonzero only for $l=k+1$. 
To evaluate $Q_{kk,lk}$, we write the pressure perturbation of the 
r-mode as 
\be
\delta p_{kk}=\rho g a\,\zeta_{k+1 k}(a)Y_{k+1 k},
\ee
where $g$ is the gravitational acceleration. The dimensionless
function $\zeta_{k+1 k}(a)$ is given by (Kokkotas \& Stergioulas 1998)
\be
\zeta_{k+1 k}(a)=\hat\Omega_s^2\left({a\over R}\right)^{k-1},
\label{eq:zeta}\ee
and the corresponding radial eigenfunction is
\be
K_{kk}(a)=i{(k+1)^2\sqrt{2k+3}\over 4k}\left({a\over R}\right)^{k-1}.
\ee
The Eulerian density perturbation is given by 
\be 
\delta\rho_{kk}=-\bxi_{kk}\cdot\nabla\rho
=a\rho S_{k+1 k}Y_{k+1 k}\delta(a-R).
\ee
But the boundary condition at the stellar surface 
gives $S_{k+1 k}(R)=\zeta_{k+1 k}(R)$, which is obtained by
requiring the Lagrangian pressure perturbation to vanish at
the surface. We then find
\be
Q_{kk,\mm+1\mm} = \rho {\hat \Oms}^2 R^{\mm+4}.
\ee
Note that this value corresponds to the normalization as defined 
by (\ref{eq:zeta}). To obtain the proper normalization
[see eq.~(\ref{eq:xinorm})], we use eq.~(\ref{eq:xir}) and find, to leading
order in $\hat\Omega_s$,
\begin{eqnarray}
\int\!d^3\!x\,\rho\,\bxi\cdot\bxi^\ast
&\simeq& k(k+1)\,\rho\int_0^R\!da\,a^4|K_{kk}(a)|^2 \nonumber\\
&=&{(k+1)^5\over 16k}\rho R^5.
\end{eqnarray}
Thus the renormalized $Q_{kk,k+1 k}$, expressed in units where 
$G=M=R=1$, is 
\be
Q_{kk,\mm+1\mm} \simeq \lb\frac{12\mm}{\pi\lp \mm+1\rp^5}\rb^{1/2}
{\hat \Oms}^2. \label{eq:qcoupr}
\ee
Thus tidal coupling will be weak for slow rotation.
With the $\bxi_\alpha$ given by (\ref{eq:xir}), we can 
similarly calculate the function $T_\alpha$ and $\varepsilon_\alpha$
[see eqs.~(\ref{eq:t}) and (\ref{eq:epsilon})], yielding
\be
\hat T_\alpha = -\mm \hat \Oms
\ee
\be
\hat \varepsilon_\alpha = \hat \omi + \mm \hat \Oms = {\hat\omr}.
\label{eq:epsilr}
\ee
The next order correction to $\hat T_\alpha$ is of order $\hat\Omega_s^5$
and can be neglected. 

Now consider resonant excitation of the $\alpha=\{kk\}$ r-mode. For 
aligned spin-orbit we require $m=k$, while to have
non-vanishing $Q_{kk,lk}$ we require $l=k+1$. Since $W_{k+1 k}=0$,
we conclude that for $\beta=0$ or $180^\circ$, 
there is no effective tidal excitation of the r-mode
(at least for the barotropic stellar model considered here).  
Resonant excitations of r-modes
can only occur for misaligned spin-orbit configurations, which we
consider in \S 4.3, using the properties of the r-mode derived above.

\setcounter{equation}{0}
\section{Non-aligned spin-orbit}

In this section we consider tidal resonances for general spin-orbit
inclination angle, $\beta$. 

\subsection{F-modes}

We again focus on the $j=k$ modes. For a given
$\alpha=\{kk\}$ mode, in addition to the $(\alpha,m)=(kk,k)$-resonance 
considered in \S 3.1, many new resonances become possible for general 
$\beta$. This can be seen by inspecting the resonance conditions
(\ref{con1})-(\ref{con3}): 
For $Q_{kk,lk}$ to be nonzero, we require $l=k,k + 2,k + 4,\cdots$
\footnote{This is because the eigenfunction of the
$\alpha=\{kk\}$ mode is even (odd) with respect to the stellar
equator for even (odd)
$k$; the tidal potential must have the same symmetry in order to 
produce nonzero coupling.}. To have nonzero tidal potential ($W_{lm}\ne 0$)
we must then have $m=k,k\pm 2,k\pm 4,\cdots$ ($m>0$). For example,
the $j=k=3$ mode can be resonantly excited at orbital frequencies
$\Omo = |\omi|$, $\Omo = |\omi| / 3$, and $\Omo = |\omi| / 5$, etc. 

The properties of the f-modes have already been calculated in 
\S 3.1. For $\omi>0$ (the spin-retrograde mode), 
the relevant Wigner ${\cal D}$-functions are
\begin{eqnarray}
{\cal D}^{(\mm)}_{\mm,\mbox{-}\mm} & = & e^{i\mm\alpha}
\lp\sin\frac{\beta}{2}\rp^{2\mm}, \label{eq:wigdkkk} \\
{\cal D}^{(\mm)}_{\mm,\mbox{-}\mm+2} & = & e^{i\mm\alpha} \sqrt{\mm(2\mm-1)}
\lp\cos\frac{\beta}{2}\rp^2 \lp\sin\frac{\beta}{2}\rp^{2\mm-2}\!\!, \\
{\cal D}^{(\mm)}_{\mm,\mbox{-}\mm+4} & = & e^{i\mm\alpha}
\sqrt{\frac{\mm(2\mm-1)(\mm-1)(2\mm-3)}{6}} \nonumber \\
&& \times \lp\cos\frac{\beta}{2}\rp^4 \lp\sin\frac{\beta}{2}\rp^{2\mm-4},
\end{eqnarray}
\begin{displaymath}
{\cal D}^{(\mm+2)}_{\mm,\mbox{-}\mm} = e^{i\mm\alpha}
\lp\sin\frac{\beta}{2}\rp^{2\mm} \lb \lp\sin\frac{\beta}{2}\rp^{4}
- 2(2\mm+2) \right.
\end{displaymath}
\be
\quad \left. \times \lp\cos\frac{\beta}{2}\rp^2 \lp\sin\frac{\beta}{2}\rp^2
 + (\mm+1)(2\mm+1)\lp\cos\frac{\beta}{2}\rp^4 \rb.
\ee
From eq.~(\ref{eq:orbchange}), we find that the orbital change
due to the $(kk,lm)$-resonance at inclination angle $\beta$
can be written as
\be
(\Delta N_{\rm orb})_{kk,lm}(\beta)=(\Delta N_{\rm orb})_{kk,lk}(180^\circ)
\times f_{kk,lm}(\beta),
\ee
where $(\Delta N_{\rm orb})_{kk,lk}(180^\circ)$ is the value for anti-aligned
spin-orbit, and the $f_{kk,lm}$ is a dimensionless function
\be
f_{kk,lm}(\beta) = \lp\frac{W_{\lr\mbox{-}\mr}}{W_{\lr\mbox{-}\mm}}\rp^2
|{\cal D}^{(\lr)}_{\mm\mbox{-}\mr}|^2 \lp\frac{\mm}{\mr}\rp^{\!\!(4\lr-11)/3}.
\ee

\begin{figure*}
\hbox{\epsffile{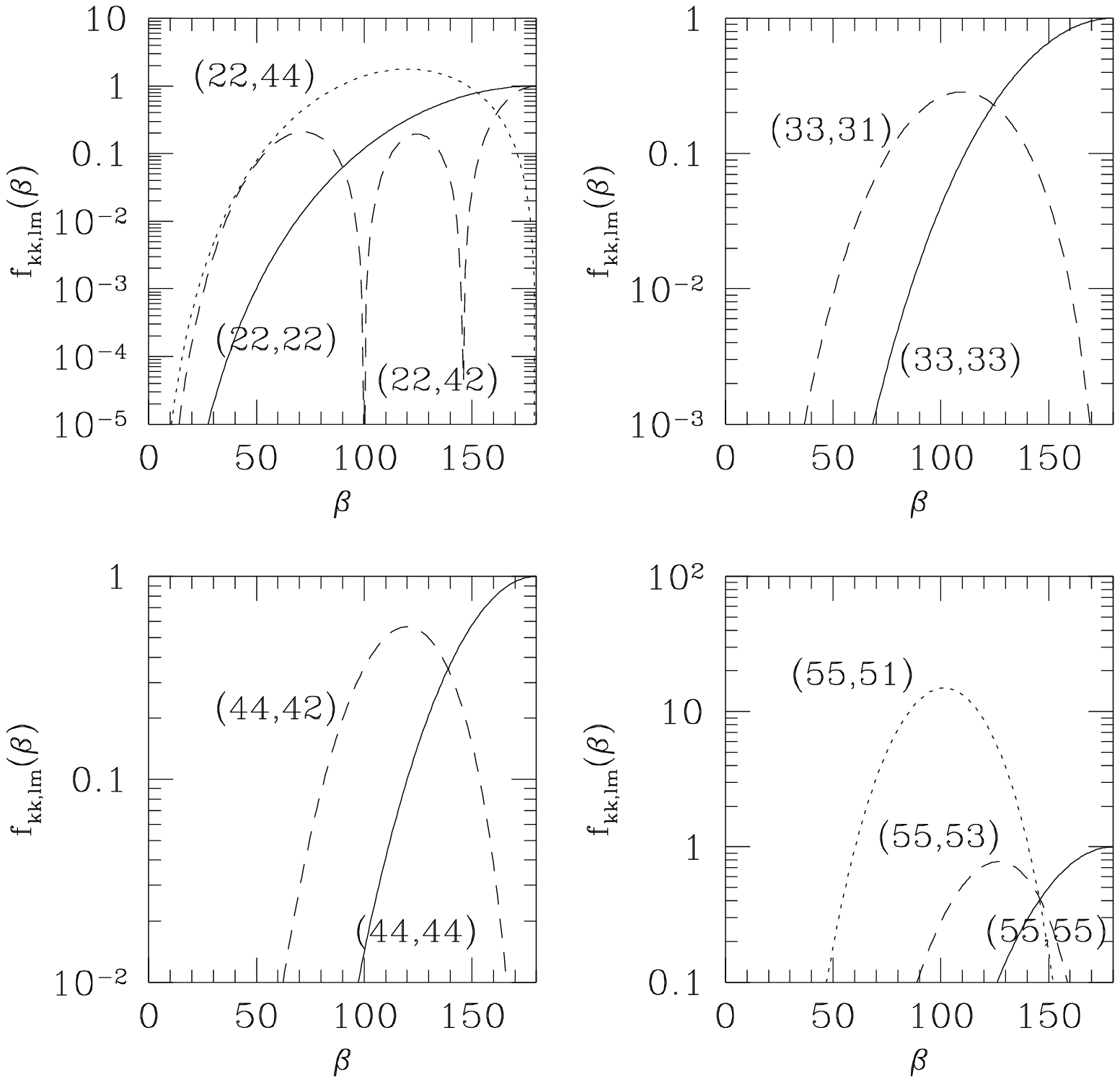}}
\caption{$f_{\mm\mm,\lr\mr}(\beta)$ as a function of spin-orbit
inclination angle, $\beta$.  $f_{\mm\mm,\lr\mr}(\beta)$ is related
to the induced orbital cycle change, $(\Delta N_{\rm orb})_{\mm\mm,\lr\mr}
(\beta)$, by $(\Delta N_{\rm orb})_{\mm\mm,\lr\mr}(\beta) =
(\Delta N_{\rm orb})_{\mm\mm,\lr\mm}(\mbox{180}^\circ) \times
f_{\mm\mm,\lr\mr}(\beta)$.  The labels in the figure give the values
of $(\lm\mm,\lr\mr)$.} \label{fig:fincline}
\end{figure*}

Figure \ref{fig:fincline} shows $f_{kk,lm}(\beta)$ as a function
of the inclination angle for a number of different resonances. 
This figure should be read together with Fig.~\ref{fig:forbchange}
to determine the actual 
orbital change at different $\nu_{\rm gw}$ and $\beta$. 
The results for the (33,53) and (33,55)-resonances are not
shown since their contributions to the orbital change are below
the limits of interest.  
It is evident from Fig.~\ref{fig:fincline} that even when the rotation
of the NS is mostly prograde with respect to the orbital motion
($\beta < 90^\circ$), there still exist modes that can be resonantly
excited and thus contribute to orbital energy dissipation.  Also of
note is the large orbital change caused by the (55,51)-resonance
relative to the (55,55) and (55,53)-resonances.  It is thus possible
for this resonance to produce an orbital change which is comparable
to those produced by the (44,4$\mr$)-resonances.

For spin-prograde modes ($\omi<0$), the above results still apply except 
that one should replace $(\Delta N_{\rm orb})_{kk,lk}(180^\circ)$ 
by $(\Delta N_{\rm orb})_{kk,lk}(0^\circ)$ and, in the expression for
$f_{kk,lm}(\beta)$, replace $\beta$ by $(180^\circ-\beta)$.

\subsection{G-modes}

The g-mode resonance considered in \S 3.2 corresponds to the
$(\alpha,lm)=(22,22)$. The function $f_{kk,lm}(\beta)$ derived
in \S 3.1 equally applies to the g-mode resonance and is plotted in 
Fig.~\ref{fig:fincline}. Although other resonances of g-modes
are also induced, their effects are below the limits of interest. 

\subsection{R-modes}

Recall that only the $\lm = \mm$ r-modes exist for barotropic stars, 
and $Q_{kk,lk} \neq 0$ only if $l=k+1$, in which case
$Q_{kk,\mm+1\mm}$ is given by (\ref{eq:qcoupr}). Resonances at
$\mr$ are restricted to $\mr = \mm+1,\mm-1,\mm-3,\ldots$ ($m>0$) 
since $W_{\mm+1\mr} = 0$ otherwise.  We consider only the
(22,3$\mr$)-resonances (for $\mr=3,1$) since the orbital effect
decreases with increasing $\lr$. The relevant Wigner ${\cal D}$-functions 
are
\begin{eqnarray}
{\cal D}^{(3)}_{23} & = & e^{2i\alpha} \sqrt{6} \,
\sin\frac{\beta}{2} \lp\cos\frac{\beta}{2}\rp^{5}, \\
{\cal D}^{(3)}_{21} & = & e^{2i\alpha} \sqrt{10} \nonumber \\
&& \!\!\!\!\! \times \,\, \sin\frac{\beta}{2} \lp\cos\frac{\beta}{2}\rp^{\!3}
\lb 2\lp\sin\frac{\beta}{2}\rp^{\!2} - \lp\cos\frac{\beta}{2}\rp^{\!2}\rb\!.
\end{eqnarray}

For the (22,33)-resonance, we find, to leading order in ${\hat\Oms}$,
that the gravitational wave frequency is given by
\be
\nu_{\rm gw}=\frac{2}{3}|\nui|\simeq \frac{8}{9}\nu_s,
\ee
where in the second equality, we have dropped the ${\hat\Oms}^3$ term
in eq.(\ref{eq:omegair}). The orbital change is given by
\begin{eqnarray}
(\Delta N_{\rm orb})_{22,33} & \simeq & +\,3.0\times10^{-4}\,
\lp\frac{Rc^2}{GM}\rp^5\frac{1}{q(1+q)^{5/3}}
\nonumber \\
&& \times \frac{|{\hat\omi}|^{1/3}}{{\hat\omr}}{\hat\Oms}^4
\lp\sin\frac{\beta}{2}\rp^{\!2}\!\lp\cos\frac{\beta}{2}\rp^{\!\!10}\!
\nonumber \\
&\simeq& +\,4.6\times10^{-5}\,M_{1.4}^{-20/3}R_{10}^{10}\frac{1}{q(1+q)^{5/3}}
\nonumber \\
&& \times \lp\frac{\nu_s}{100 \mbox{Hz}}\rp^{\!\!10/3}\!\!
\lp\sin\frac{\beta}{2}\rp^{\!2}\!\lp\cos\frac{\beta}{2}\rp^{\!\!10}\!.
\label{del2233}
\end{eqnarray}
Recall that the r-mode has negative energy, and as a result,
$(\Delta N_{\rm orb})_{\alpha,\lr\mr}$
is positive, i.e., the resonance slows down the inspiral.
At the maximum $\nu_{\rm gw}=1000$~Hz (or $\nu_s\simeq 1125$~Hz) and
$\beta=48.2^\circ$ (which would give the maximum orbital change), we have
\be
(\Delta N_{\rm orb})_{22,33}^{\rm (max)}\simeq 0.010\,
M_{1.4}^{-20/3}R_{10}^{10}\frac{1}{q(1+q)^{5/3}}.
\label{delmax}\ee
The orbital change increases with increasing spin frequency.
But note that our analysis of r-mode is based on the
assumption of slow rotation, thus (\ref{delmax}) should be
considered as an estimate. 

For the (22,31)-resonance, we have
\be
\nu_{\rm gw}=2|\nui|\simeq \frac{8}{3}\nu_s
\ee
and
\begin{displaymath}
(\Delta N_{\rm orb})_{22,31} \simeq +\,4.3\times10^{-4}\,
\!\lp\frac{Rc^2}{GM}\rp^{\!\!5}\!\!\frac{1}{q(1+q)^{5/3}}
\frac{|{\hat\omi}|^{1/3}}{{\hat\omr}}{\hat\Oms}^4
\end{displaymath}
\begin{displaymath}
\quad \qquad \times
\lp\sin\frac{\beta}{2}\rp^{\!2} \lp\cos\frac{\beta}{2}\rp^{\!6}
\lb 2\lp\sin\frac{\beta}{2}\rp^{\!2}
- \lp\cos\frac{\beta}{2}\rp^{\!2}\rb^2\!\!
\end{displaymath}
\begin{displaymath}
\quad \qquad \simeq +\,6.6\times10^{-5}\,M_{1.4}^{-20/3}R_{10}^{10}
\frac{1}{q(1+q)^{5/3}} \lp\frac{\nu_s}{100 \mbox{Hz}}\rp^{\!\!10/3}\!\!
\end{displaymath}
\be
\quad \qquad \times
\lp\sin\frac{\beta}{2}\rp^{\!2} \lp\cos\frac{\beta}{2}\rp^{\!6}
\lb 2\lp\sin\frac{\beta}{2}\rp^{\!2}
- \lp\cos\frac{\beta}{2}\rp^{\!2}\rb^2\!\!.
\ee
Again, at the optimal parameters $\nu_{\rm gw}=1000$~Hz 
(or $\nu_s\simeq 375$~Hz) and $\beta=34.4^\circ$, we have 
\be
(\Delta N_{\rm orb})_{22,31}^{\rm (max)}\simeq 2.0\times 10^{-4}\,
M_{1.4}^{-20/3}R_{10}^{10}\frac{1}{q(1+q)^{5/3}}.
\label{delmax2}\ee

Thus we conclude that for canonical NS parameters, 
the (22,33) and (22,31)-resonances (and indeed all r-mode resonances) produce
orbital changes that are below the limits of interest. However,
if we consider a larger NS radius, such as $R=15$~km for
a $M=1.4M_\odot$ NS, which is allowed for some nuclear equations of state 
(e.g, Shapiro \& Teukolsky 1983), then the orbital change
for the $(22,33)$-resonance may become significant
($\Delta N_{\rm orb} \sim 0.6$).

\section{Summary and Discussion}

In this paper, we have presented a systematic study on
the resonant mode excitations of rotating neutron stars in coalescing
binaries and their effects on the gravitational waveform.
The essential effect of rotation is that it can modify the mode 
frequency in the inertial frame, thereby changing the strength of 
the g-mode resonance, which already exists for nonrotating neutron stars,
and also making a variety of new resonances involving f-modes and r-modes
possible. 

We find that for the f-mode resonance to occur during the
last few minutes of the binary inspiral, with the emitted gravitational
wave frequency $10~\mbox{Hz}<\nu_{\rm gw}<1000~\mbox{Hz}$
(the sensitivity band of the 
interferometric gravitational wave detector), the neutron star
must have very rapid rotation (see Fig.~\ref{fig:fspin}), e.g.,
assuming canonical neutron star mass and radius, $M=1.4M_\odot$ and
$R=10$~km, respectively, $\nu_s\go 710$~Hz
for the $(jk,m)=(22,2)$-resonance (see the beginning of \S 2.2 for notation)
and
$\nu_s\go 570$~Hz for the $(33,3)$-resonance.
Observations (e.g., PSR 1913+16 has spin period $59$~ms, 
and PSR 1534+12 has $38$~ms) and our current understanding of 
the formation of neutron star binaries seem to indicate  
that neutron stars in compact binaries (with
another neutron star or a black hole as a companion) can never achieve
such a rapid rotation, although such a rotation rate is physically
allowed.  For a $M=1.4M_\odot$ and $R=15$~km neutron star,
the critical frequencies are lowered, e.g.,
$\nu_s\go 330$~Hz for the $(22,2)$-resonance and
$\nu_s\go 260$~Hz for the $(33,3)$-resonance (see Fig.~\ref{fig:fspin15}).
If the neutron star in a coalescing binary indeed has the required
rapid rotation, then the induced orbital change due to the f-mode resonance
is significant (see Fig.~\ref{fig:forbchange}
and Fig.~\ref{fig:forbchange15}) and must be included in the 
templates of waveforms used for searching gravitational wave signals.

The resonance of g-modes is strongly affected by even a modest
rotation (see Fig.~\ref{fig:gspin}). Although there is large uncertainties 
associated with the property of the g-mode (since it depends on the
symmetry energy of nuclear matter), we show that for 
a canonical $1.4M_\odot$ neutron star, the orbital change
$\Delta N_{\rm orb}$ due to the g-mode resonance is below the
$10^{-2}$ level (see Fig.~\ref{fig:gorbchange}). Such a small
phase error in the waveform
is unimportant for the search of gravitational wave signals from the binary
(E. Flanagan 1998, private communication). However, we note that
if we consider very low-mass ($M\lo 0.5M_\odot$) neutron stars, which 
have large $R/M$ ratio, the orbital change due to g-mode resonance may well be
significant. Of course, although such low-mass neutron stars are physically
allowed, there is no astrophysical evidence for their existence. 

Resonant excitations of r-modes occur for misaligned spin-orbit 
configurations. Since the r-mode always has negative energy, the resonance
actually transfers energy to the orbit, slowing down the inspiral. 
Since the tidal coupling of the r-mode depends strongly on the
stellar rotation rate, we find that for canonical neutron star mass and
radius ($1.4M_\odot$ and $10$~km, respectively),
the orbital change $\Delta N_{\rm orb}$
is negligible. However, if we consider a larger stellar radius,
such as $15$~km for a $1.4M_\odot$ neutron star (as is allowed by some
nuclear equations of state), then $\Delta N_{\rm orb}$ may become
important, especially in the high frequency band ($\nu_{\rm gw}\sim 1000$~Hz)
[see eqs.~(\ref{del2233}) and (\ref{delmax})].

Finally, we note that our calculations of the properties of the f-modes
and r-modes are based on incompressible neutron star models. 
Our results for the r-modes should break down for very high rotation rates. 
Since neutron stars are not exactly barotropic, other r-modes,
in addition to the $j=k$ mode considered in this paper, can also exist. 
It is desirable to study the resonance effects using more realistic 
neutron star models.  

\section*{Acknowledgments}

We thank Eanna Flanagan for useful discussion.  We are also
grateful to Scott Hughes for helpful comments and suggestions.
D.L. is supported by a Alfred P. Sloan Foundation fellowship.


\label{lastpage}

\clearpage

\clearpage
 
\clearpage
 
\clearpage

\clearpage

\clearpage


\begin{thebibliography}{99}

\bibitem{} Abramovici, A. 1992, Science, 256, 325

\bibitem{} Andersson, N. 1998, ApJ, 502, 708

\bibitem{} Andersson, N., Kokkotas, K.D., Schutz, B.F. 1999, ApJ, 510, 846

\bibitem{} Blanchet, L. et al. 1995, Phys. Rev. Lett., 74, 3515

\bibitem{} Baumgarte, T.~W. et al.~1997, Phys. Rev. Lett., 79, 1182

\bibitem{} Bildsten, L., Cutler, C. 1992, ApJ, 400, 175

\bibitem{} Bildsten, L., Cutler, C. 1995, ApJ, 449, 800

\bibitem{} Bradaschia, C. et al. 1990, NIM, A289, 518

\bibitem{} Brady, P.~R., Creighton, J.~D.~E., Thorne, K.~S.
1998, Phys. Rev. D58, 61501

\bibitem{} Bryan, G.H. 1889, Phil. Trans. Roy. Soc. London, A180, 187

\bibitem{} Chandrasekhar, S. 1969, Ellipsoidal Figures of Equilibrium
(Yale University Press)

\bibitem{} Chandrasekhar, S. 1970, Phys. Rev. Lett., 24, 611

\bibitem{} Comins, N. 1979, MNRAS, 189, 255

\bibitem{} Cutler, C. et al. 1993, Phys. Rev. Lett., 70, 2984

\bibitem{} Cutler, C., Flanagan, E.~E. 1994, Phys. Rev. D49, 2658

\bibitem{} Davies, M.~B., et al. 1994, ApJ, 431, 742

\bibitem{} Finn, L.S. 1987, MNRAS, 227, 265

\bibitem{} Friedman, J.L., Morsink, S.M. 1998, ApJ, 502, 714

\bibitem{} Friedman, J.L., Schutz, B.F. 1978a, ApJ, 221, 937

\bibitem{} Friedman, J.L., Schutz, B.F. 1978b, ApJ, 222, 281

\bibitem{} Kochanek, C.S. 1992, ApJ, 398, 234

\bibitem{} Kokkotas, K.D., Stergioulas, N. 1999, A\&A, 341, 110

\bibitem{} Lai, D. 1994, MNRAS, 270, 611

\bibitem{} Lai, D. 1997, ApJ, 490, 847

\bibitem{} Lai, D. 1998, MNRAS, submitted (astro-ph/9806378)

\bibitem{} Lai, D., Rasio, F.~A., Shapiro, S.~L. 1994, ApJ, 420, 811

\bibitem{} Lai, D., Shapiro, S.~L. 1995, ApJ, 443, 705

\bibitem{} Lai, D., Wiseman, A.~G. 1996, Phys. Rev. D54, 3958

\bibitem{} Lindblom, L. 1995, ApJ, 438, 265

\bibitem{} Lindblom, L., Owen, B.J., Morsink, S.M. 1998, Phys. Rev. Lett.,
80, 4843

\bibitem{} McDermott, P.N., Van Horn, H.M., Hansen, C.J. 1988, ApJ, 325, 725

\bibitem{} New, K.~C.~B., Tohline, J.~E. 1997, ApJ, 490, 311

\bibitem{} Press, W.H., Teukolsky, S.A. 1977, ApJ, 213, 183

\bibitem{} Provost, J., Berthomieu, G., Rocca, A. 1981, A\&A, 94, 126

\bibitem{} Rasio, F.A., Shapiro, S.L. 1995, ApJ, 438, 887

\bibitem{} Reisenegger, A., Goldreich, P. 1992, ApJ, 395, 240

\bibitem{} Reisenegger, A., Goldreich, P. 1994, ApJ, 426, 688

\bibitem{} Ruffert, M., Janka, H.~T., Takahashi, K., Schafer, G.
1997, A\&A, 319, 122

\bibitem{} Saio, H. 1982, ApJ, 256, 717

\bibitem{} Shapiro, S.L., Teukolsky, S.A. 1983, Black Holes, White
Dwarfs, and Neutron Stars (John Wiley \& Sons, Inc.)

\bibitem{} Shibata, M., Nakamura, T., Oohara, K. 1992, Prog. Theor. Phys.,
88, 1079

\bibitem{} Teukolsky, S.A. 1998, in Wald, R.M., ed, Black Holes and
Relativistic Stars (University of Chicago Press)

\bibitem{} Thorne, K.S. 1987, in Hawking, S.W., Israel, W., eds,
300 Years of Gravitation (Cambridge University Press)

\bibitem{} Thorne, K.S. 1998, in Wald, R.M., ed, Black Holes and
Relativistic Stars (University of Chicago Press)

\bibitem{} Uryu, K., Eriguchi, Y. 1998, MNRAS, 296, L1

\bibitem{} Wybourne, B.~G. 1974, Classical Groups for Physicists
(New York: John Wiley)

\bibitem{} Zhuge, X., Centrella, J.M., McMillan, S.L.W. 1996, Phys. Rev. D54,
7261

\end{thebibliography}
\end{document}